\documentclass[12pt]{article}
\usepackage{epstopdf, graphicx}

\usepackage{times}
\usepackage{epstopdf, graphicx}
\usepackage{amssymb,amsmath}
\usepackage{scicite}

\newcommand\ion[2]{#1$\;${\small\rmfamily{#2}}\relax}%
\def\tll{$\tau_{\rm LL}$}
\def\mtll{\tau_{\rm LL}}

\def\sci#1{{\; \times \; 10^{#1}}}

\def \kms  {km~s$^{-1}$}
\def \mkms  {{\rm km~s^{-1}}}

\def \lya  {Ly$\alpha$}

\def \lyb  {Ly$\beta$}

\def \ly5  {Ly-5}
\def \ly6  {Ly-6}
\def \ly7  {Ly-7}
\def \obary  {$\Omega_{b,0}$}
\def \mobary  {\Omega_{b,0}}
\def \mzsun  {Z_\odot}
\def \mnh  {n_{\rm H}}
\def \nh  {$n_{\rm H}$}
\def \mvhi  {n_{\rm HI}}

\def \mxhi  {x_{\rm HI}}
\def \nhi  {$N_{\rm HI}$}
\def \mnhi  {N_{\rm HI}}
\def \lnhi {$\log N_{\rm HI}$}

\newcommand{\cm}[1]{\, {\rm cm^{#1}}}

\def\zsun {Z_\odot}

\def \zllsone    {$z=3.410883$}       
\def \tabzone    {$3.41088$}          
\def \dhone      {$-4.69 \pm 0.13$}   
\def \nhione     {$17.95 \pm 0.05$}   
\def \metone     {$<-4.2$}            
\def \zsunone    {$Z<10^{-4.2}\zsun$}
\def \nonelong   {SDSS J113418.96+574204.6} 
\def \nameone    {J1134+5742}               

\def \zllstwo    {$z=3.096221$}       
\def \zsuntwo    {$Z<10^{-3.8}\zsun$}
\def \nametwo    {Q$0956+122$}        

\def\llso{LLS1134a}
\def\llsta{LLS0956A}
\def\llstb{LLS0956B}

\def\figmetsum{1}
\def\figllso{2}
\def\figllsta{3}
\def\figdh{4}
\def\figpreenr{5}
\def\tablssprop{1}

\def\figlris{S1}
\def\figlymano{S2}
\def\figlymant{S3}
\def\figmetline{S4}
\def\figcla{S5}
\def\figudis{S6}

\def\tabfitllso{S1}
\def\tabmetlim{S2}
\def\tabdeutsys{S3}

\newenvironment{sciabstract}{%
\begin{quote} \bf}
{\end{quote}}

\newcounter{lastnote}
\newenvironment{scilastnote}{%
\setcounter{lastnote}{\value{enumiv}}%
\addtocounter{lastnote}{+1}%
\begin{list}%
{\arabic{lastnote}.}
{\setlength{\leftmargin}{.22in}}
{\setlength{\labelsep}{.5em}}}
{\end{list}}

\title{Detection of Pristine Gas Two Billion Years after the Big Bang}
\date{}

\author{
Michele Fumagalli$^{1\ast}$,
John M. O'Meara$^{2}$, and
J. Xavier Prochaska$^{3}$
\\
\normalsize{$^{1}$Department of Astronomy and Astrophysics, UC Santa Cruz, CA}\\
\normalsize{$^{2}$Department of Chemistry and Physics, Saint Michael's College, Colchester, VT}\\
\normalsize{$^{3}$University of California Observatories-Lick Observatory, UC Santa Cruz, CA}\\
\\
\normalsize{$^\ast$To whom correspondence should be addressed; E-mail: mfumagalli@ucolick.org}
}

\begin{document}

\maketitle
\begin{sciabstract}  
In the current cosmological model, only the three lightest elements were created in the first 
few minutes after the Big Bang; all other elements were produced later in stars.  
To date, however, heavy elements have been observed in all astrophysical environments.  
We report the detection of two gas clouds with no
discernible elements heavier than hydrogen. These systems exhibit the lowest heavy-element
abundance in the early universe and thus are potential fuel for the most metal poor halo stars.
The detection of deuterium in one system at the level predicted by primordial nucleosynthesis provides 
a direct confirmation of the standard cosmological model. The composition of 
these clouds further implies that the transport of heavy elements from galaxies to their 
surroundings is highly inhomogeneous.
\end{sciabstract}

In modern cosmological theory, the light elements and their
isotopes are produced during the first few minutes after the Big Bang when the universe cools
during expansion from temperatures $T \sim 10^9$\,K to below $\sim 4\times 10^8$\,K.
In this brief epoch, termed Big Bang Nucleosynthesis (BBN),   
D, $^3$He, $^4$He, and $^7$Li are synthesized with an abundance 
ratio relative to hydrogen that is sensitive to the cosmic 
density of ordinary matter (i.e. the baryon density \obary). 
BBN theory also predicts negligible production of the heavy elements 
with abundance ratios $\rm X/H < 10^{-10}$ and one must await the 
physical conditions that are typical of the  
stellar interiors \cite{bur57}. 

The analysis of gas observed in absorption along the lines-of-sight to 
high-redshift quasars, distant galaxies that host supermassive black holes, 
is a powerful probe of the BBN yields.  Particular attention has been given 
to deuterium, partly due to observational convenience but also because the
$\rm D/H$ abundance ratio is very sensitive to \obary. For quasar sight lines, 
the measured $\log \rm (D/H)  = -4.55 \pm 0.03$ \cite{ome06,pet08b}
translates into $\mobary h^2(\rm BBN) = 0.0213 \pm 0.0010$, 
which is fully consistent with the value inferred from the Cosmic Microwave 
Background (CMB) power spectrum 
$\mobary h^2(\rm CMB) = 0.02249^{+0.00056}_{-0.00057}$ (WMAP7 \cite{kom11}).
This excellent agreement between two essentially independent
experiments stands as a marked triumph of the Big Bang theory.

On the other hand, all of the systems with measured D
have heavy-element abundances that exceed, by many orders of
magnitude, the BBN prediction. In fact, despite measurement of
thousands of galaxies from the early universe (e.g.\ \cite{pro03,erb06}) 
and careful study of the diffuse gas that permeates the universe 
(e.g.\ \cite{sch03,sim04}), one has yet to detect 
anything near primordial enrichment.  For structures denser than the mean cosmic 
density (Fig. \figmetsum), the high-redshift universe has
exhibited a floor in the metallicity $Z$, 
the mass fraction of elements heavier than hydrogen,
at $\sim 1/1000$ of the solar abundance ($\zsun$). 
Similarly,  among several old and iron-poor
stars, only one has a metallicity  $Z\sim 10^{-4} \zsun$ \cite{caf11}, 
with all the remaining having enhanced C or O abundances.
The existence of a minimum level of enrichment at about $1/1000$ solar
has been associated with the metal production in Population III (PopIII) stars, 
primordial stars that form in metal free environments via H$_2$ cooling.
In fact, models and numerical simulations (e.g.\ \cite{mac03,wis08}) show that 
ejecta from this first stellar population can enrich the interstellar medium (ISM) 
of the host halos up to $\sim 10^{-3} \zsun$ and pollute 
the surrounding intergalactic medium (IGM) as soon as 1 billion years
after the Big Bang ($z \sim  6$).

In this work, we report on the hydrogen and metal properties of two gas 
`clouds' at $z \sim3$, when the universe was only two billion years old. 
We observed quasars \nonelong\ ($z_{\rm em} = 3.522$) 
and \nametwo\ ($z_{\rm em} =3.297$) on UT January 3 and 5, 2006
and UT April 7, 2006 with the HIRES spectrometer 
on the Keck\,I telescope on Mauna Kea (see SOM text 1).
Previous low-resolution spectra of these quasars from the Sloan Digital Sky Survey
had shown significant absorption at wavelengths $\lambda < 4000$\AA\ 
characteristic of substantial optical depth \tll\ at the \ion{H}{I} 
Lyman limit (at wavelength $\lambda < 912$\AA), typical of the 
Lyman limit systems (LLSs). 

Our Keck/HIRES spectrum of \nameone\ reveals a sharp break in the flux at $\sim 4000$ \AA, 
indicating the presence of a LLS at $z\sim 3.4$
with $\mtll > 2$ (hereafter named LLS1134; Fig. \figllso), 
confirming the lower resolution data. 
A search for absorption lines at a redshift 
consistent with the Lyman break further reveals the presence of the \ion{H}{I}
Lyman series through to Lyman--22 (\ion{H}{I}~913.5), corresponding to $\lambda \sim
4030$\AA.  In the high resolution spectrum, 
two distinct absorbers can be identified within LLS1134, the main system
(hereafter \llso) at \zllsone\ and a weaker component  (named LLS1134b)
at $z=3.41167$, separated by $\delta v \sim 54 ~\mkms$. 
In contrast, two flux decrements in the normalized HIRES 
spectrum of \nametwo\ (Fig. \figllsta) are visible at $\sim 3860$ \AA\ 
and $\sim 3750$ \AA, revealing the presence of two
LLSs with $\mtll \sim 1$ at $z\sim3.22$ (\llsta) and $z\sim3.10$ (\llstb).

A closer inspection of these spectra also reveals
no detectable metal-line absorption at the velocity of the
strongest \ion{H}{I} component for neither \llso\ nor \llstb.
Such a complete absence of heavy element absorption has not 
been previously reported for data with comparable sensitivity.    
For \llso,  besides the \ion{H}{I} Lyman series, the only other detected transitions 
are \ion{D}{I} \lya\ and \lyb, offset by the appropriate 
$\delta v =-82~\mkms$ from the hydrogen absorption.

To characterize the physical properties of these LLSs (summarized in Table~\tablssprop), 
we modeled the hydrogen and deuterium absorptions using a $\chi^2$ minimization algorithm
(Table \tabfitllso; Fig. \figlymano\ and \figlymant, see SOM text 2)
and we derived upper limits on the column densities of various ionization states 
of heavy elements (Table~\tabmetlim; Fig. ~\figmetline, see SOM text 3). 
To translate these limits into constraints on the gas metallicity, 
it is necessary to make assumptions on the
ionization state of the gas (see SOM text 4).  
Every LLS with $\mnhi \lesssim 10^{19}
\cm{-2}$ analyzed to date has exhibited absorption characteristic of a
predominantly ionized gas (e.g.\ \cite{prt10,pro99}).
Standard interpretation is that the medium has been photoionized by an
external radiation field, presumably a combination of the
extragalactic UV background (EUVB) generated by the cosmological
population of quasars and galaxies together with emission from local sources
(e.g.\ a nearby galaxy).  This conclusion is based on comparison of
the observed ionic column densities of heavy elements with simple
photoionization models.  Every previous LLS has shown substantial
absorption from doubly and triply ionized species, e.g. Si$^{++}$ and
C$^{+3}$, which trace predominantly ionized gas.
Parametrizing the ionization state in terms of the ionization
parameter $U \equiv \Phi/(c \, \mnh)$, with $\Phi$ the flux of ionizing
photons and \nh\ the gas volume density,  all previously analyzed LLSs
have exhibited $U \geq 10^{-3}$ corresponding to $n_{\rm H} \leq 10^{-2}
\cm{-3}$ at $z\sim 3$ (Fig. \figudis).

Adopting this $U$ value as a limit to the ionization state
and current estimates for the spectral shape of the EUVB \cite{haa11}, 
we infer metallicities of \zsunone\ and \zsuntwo\ for 
\llso\ and \llstb\ (see Fig. \figcla).  
These upper limits are 100 to 1000 times lower than typical
measurements of LLSs and over an order of magnitude lower
than any previous observed metallicity at $z>2$ (Fig. \figmetsum).
These limits are only comparable to the abundances detected in the most 
metal poor star \cite{caf11} and are suggestive of a primordial 
composition. 
Super-solar metallicity is commonly found in the surroundings of quasars 
as well as in a few LLSs. Metallicities between solar and $\sim 1/10$ solar 
are typical of galaxies, and sub-solar metal enrichment down to $10^{-3} \zsun$ 
is characteristic of the ISM and IGM at early epochs. 
Remarkably, the most iron poor stars in the Galactic halo, thought to be the repository of
the first generation of metals, have total heavy element abundance 
comparable to these limits, but generally above $\sim 1/1000$ solar. 
Our analysis uncovers regions of the universe at $z<6$ with essentially primordial enrichment,
whose traces can be found within the oldest stellar populations in the present universe \cite{caf11}.

The detection of deuterium in the metal free \llso\
provides a direct confirmation of the BBN.
From the analysis of the D and H absorption lines, 
we derive $\log \rm (D/H) = -4.69 \pm 0.13$ (see SOM text 2). 
The observed value is consistent at $1\sigma$ with 
the theoretical predicted value $\log \rm (D/H) =-4.592$ \cite{ste06}, 
assuming $\mobary$ from WMAP7. 
This measurement is in agreement with previous determinations in quasar absorption 
line systems at $z>2$ (Table \tabdeutsys; Fig. \figdh). 
Because the lack of metals in \llso\ confirms its pristine
composition, this agreement strengthens the hypothesis that at low
metallicities ($10^{-2} \mzsun$ or less) the observed 
deuterium abundances are representative of the primordial value
\cite{kir03} and astration cannot be responsible for the
lingering scatter in the observed $\rm D/H$. Deuterium 
abundances from quasar absorption line systems 
are therefore solid anchor points for models 
of galactic chemical evolution (e.g. \cite{rom06}).
Combining $\rm D/H$ in \llso\ with values from the known D-bearing systems, 
we obtain a logarithmic weighted mean $\overline{\log \rm (D/H)} = -4.556 \pm 0.034$ that 
translates into  $\Omega_{b,0} h^2(\rm BBN) = 0.0213 \pm 0.0012$,  
after accounting for both random and systematic errors \cite{pet08b}.
Consistent with previous studies, we do not include in the weighted mean the error on the
assumed level of the quasar continuum light. 

The absence of metals in \llso\ and \llstb\ is outstanding
also in the framework of theories for the metal enrichment of cosmic structures 
(Fig. \figpreenr). Numerical simulations suggest that LLSs typically arise 
in galaxies (e.g. \cite{koh07}), 
in dense gas above $\sim 100$ times the mean cosmic density $\rho_{\rm mean}$. 
At the same time, in models of the IGM enrichment, metals are ejected to 
hundreds of kpc from star forming regions, resulting in 
substantial pollution of the nearby gas. As a consequence, the metallicity predicted for 
stars or for the ISM at $z<4$ \cite{opp11} ranges between $0.1-1 \zsun$, 
three orders of magnitude higher than the limits inferred for these two LLSs. 
Similarly, the metallicity predicted for the hot halo of galaxies and for the 
surrounding IGM \cite{opp11,her03} exceeds by a factor of 10 or more
the limits for \llso\ and \llstb. Contrary to any prediction and any previous
observation, these two LLSs reside at significant overdensity ($\rho/\rho_{\rm mean}<850$)
but in an unpolluted portion of the universe.

Metallicity below $10^{-4} \zsun$ appears even exceptional when 
compared to the level of pre-enrichment from PopIII stars that is 
predicted by models \cite{mac03,yos04,wis10} already 500 million years 
after the Big Bang ($z\sim 10$). Our limits place additional constraints on
the widespread dispersal of metals from primordial stellar populations and the first
generations of galaxies. Remarkably,
the gas we detected could in principle fuel PopIII star-formation at $z \sim 3$, because
its metallicity  lies at or even below 
the minimum enrichment required for metal cooling to induce fragmentation in the 
collapsing material in the absence of dust \cite{bro01}.  
Given a lower limit on the LLS physical size
$\ell = N_{\rm HI}/(x_{\rm HI} n_{\rm H})$, with $x_{\rm HI}$ the neutral fraction and
$n_{\rm H}$ the total hydrogen volume density, we can infer the total 
hydrogen mass in these clouds $M_{\rm H} = m_{\rm p} N_{\rm H} \ell^2$.
We find $M_{\rm H} \gtrsim 6.2 \times 10^6~\rm M_\odot$ for \llso\ and  
$M_{\rm H} \gtrsim 4.2 \times 10^5~\rm M_\odot$  for \llstb,
comparable to the mass of the mini-halos 
where the first generation of stars formed (e.g. \cite{wis10}). Therefore, if this gas were able to 
collapse further and shield from the ambient UV radiation, it would 
potentially give rise to PopIII stars two billion years after the transition between 
the first and second generation of stars (PopII) is thought to have occurred.
Thus, pair-production supernovae associated with the death of these massive and metal-free 
stars may be found even at modest redshifts \cite{sca05}.

The pristine composition of \llso\ and \llstb\ can be reconciled with model predictions 
and previous observations if mixing of metals within the IGM is an 
inefficient and inhomogeneous process.
A varying degree of metal enrichment is seen in multiple components of LLSs 
\cite{prt10,pro99,dod01}, implying that mixing does not operate 
effectively on small scales. The detection of ionized metals in LLS1134b, 
the weaker component at $+54~ \rm km~s^{-1}$ from \llso, reinforces this point. 
Further, studies of metal systems in the low density and diffuse IGM
suggest that at least some of the ionized heavy elements (e.g. \ion{C}{IV}) are in 
small and short-lived clumps \cite{sch07}. A low volume filling factor for 
metals is also consistent with theories of metal ejection
from supernovae, in which most of the heavy elements are initially confined in small 
bubbles \cite{fer00} and only subsequently diffuse in the surrounding IGM.
Plausibly, \llso\ and \llstb\ originate in a filament of the cosmic 
web where primordial regions coexist with enriched pockets of gas.
If the metal enrichment is highly inhomogeneous, these two LLSs could just be 
the tip of the iceberg of a much larger population of unpolluted absorbers that 
trace a large fraction of the dense IGM.

Beside the implications for the BBN and the metal distribution, the detection of 
metal free LLSs is tantalizing in the context of galaxy formation and evolution. 
Modern theory and simulations predict that most of the gas that sustains star formation 
is accreted in galaxies through dense and narrow streams, known as 
cold flows \cite{ker05,dek09}. These gaseous filaments are highly 
ionized by both the EUVB and the radiation escaping from the central star 
forming regions. Cold flows should therefore appear as LLSs
in the spectra of bright quasars \cite{fau11,fum11,van11}. 
Further, in contrast to metal enriched gas that is outflowing from galaxies, 
this infalling material is expected to be metal poor \cite{fum11}. 
Although direct observational evidence of cold flows is still lacking, 
these streams are thought to be ubiquitous at high-redshift and primordial LLSs such as 
\llso\ and \llstb\ are ideal candidates for this elusive mode of accretion.

\bibliographystyle{Science}
\bibliography{primordial_biblio}

\begin{scilastnote}
\item We thank A. Aguirre, N. Lehner, and P. Madau for providing comments on this manuscript. 
We would like to thank the MPIA at Heidelberg for their hospitality. J.X.P. acknowledges support 
from the Humboldt Foundation. Support for this work came from NSF grant AST0548180. 
We acknowledge the use of VPFIT. This work is based on observations made at the W. M. Keck 
Observatory, which is operated as a scientific partnership among the California 
Institute of Technology, the University of California and NASA. The Observatory was 
made possible by the generous financial support of the W. M. Keck Foundation. 
The authors wish to recognize and acknowledge the very significant cultural role and
reverence that the summit of Mauna Kea has always had within the
indigenous Hawaiian community.  We are most fortunate to have the
opportunity to conduct observations from this mountain. 
The data reported in this paper are available through the Keck Observatory Archive (KOA).
\end{scilastnote}

\vskip 1cm
\noindent
{\bf Supporting Online Material}\\
www.sciencemag.org\\
Supporting text\\
Tables S1-S3\\
Figures S1-S6\\
References (35-68)
\clearpage

\clearpage
\begin{tabular*}{\textwidth}{@{\extracolsep{\fill}}lcc}
\hline
\hline
&\llso&\llstb\\
\hline
Redshift                      &$3.410883\pm0.000004$                & $3.096221\pm0.000009$  \\
$\log N_{\rm HI}$             &$17.95\pm0.05$                       & $17.18\pm0.04$         \\
$\log \rm D/H$                &$-4.69\pm0.13^{a}$                   &  -                     \\
$b_{\rm HI}$ (km s$^{-1}$)    &$15.4\pm0.3$                         & $20.2\pm0.8$           \\
Temperature (K)               &$<(1.43 \pm 0.05) \times 10^4$       & $<(2.48 \pm 0.19) \times 10^4$      \\
Metallicity ($Z_{\odot}$)     &$<10^{-4.2}$                         & $<10^{-3.8}$             \\
$\log x_{\rm HI}$             &$<-2.10$                             & $<-2.40$                \\
$\log n_{\rm H}$              &$<-1.86$                             & $<-1.98$                \\
$\log U^b$                    &$>-3$                                & $>-3$                   \\
\hline 
\hline
\end{tabular*}
{\footnotesize $^a$ Including \llso, the current best estimate for the primordial 
deuterium abundance becomes $\overline{\log \rm (D/H)} = -4.556 \pm 0.034$. $^b$ 
The listed values are physically motivated, but not directly measured. Note that
the metallicity, $x_{\rm HI}$, and $n_{\rm H}$ depend on the assumed value.}
\newline
\newline

\noindent
{\bf Table \tablssprop:} 
{\small Summary of the physical properties for \llso\ and \llstb. For each system, we 
present: redshift, the hydrogen column density ($N_{\rm HI}$), the deuterium 
 abundance ($\rm D/H$), the Doppler parameter ($b_{\rm HI}$), the temperature, 
 the metallicity, the hydrogen neutral fraction ($x_{\rm HI}$), the total 
 hydrogen volume density ($n_{\rm H}$), and the ionization parameter ($U$).}

\clearpage
\begin{figure*}[!ht]
\begin{center}
\includegraphics[scale=0.6,angle=90]{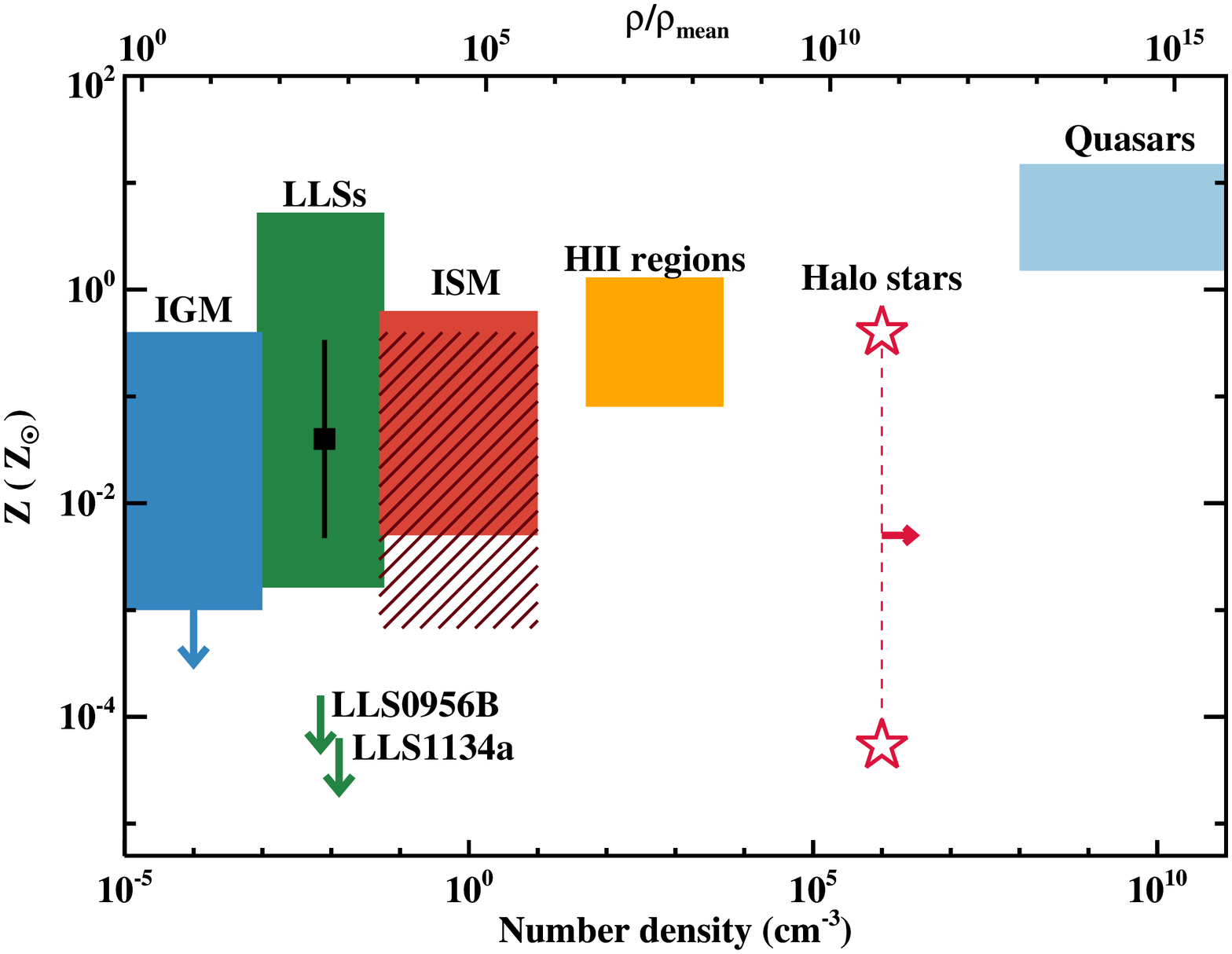}
\end{center}
\end{figure*}
\noindent
{\bf Figure \figmetsum:}
{\small 
Peak to peak variation of the observed metallicity 
in $z\gtrsim 2$ cosmic structures at different densities.
The blue, green and red rectangles show the spread in observed
metallicities for diffuse gas in the universe; respectively, these are
the IGM, LLSs, and galactic ISM.
Orange rectangle: \ion{H}{II} regions in galaxies.
Light blue rectangle: quasar broad line regions. The black point with error bars marks 
the mean metallicity and the standard deviation for $z>1.5$ LLSs.
Galactic halo stars (stars connected with a dashed line) are represented at arbitrary density
for visualization purposes. The top axis translates the number density in the 
overdensity above the mean baryon cosmic density at $z=3.5$.
The upper limits on the metallicity for \llso\ and \llstb\ 
are shown with green arrows, assuming $\log U=-3$.
Higher ionization parameters would shift these limits to lower densities 
and lower metallicity. See the SOM text 5 for additional details on the 
observations presented in this figure.}

\clearpage
\begin{figure*}[!ht]
\begin{center}
\includegraphics[scale=0.58]{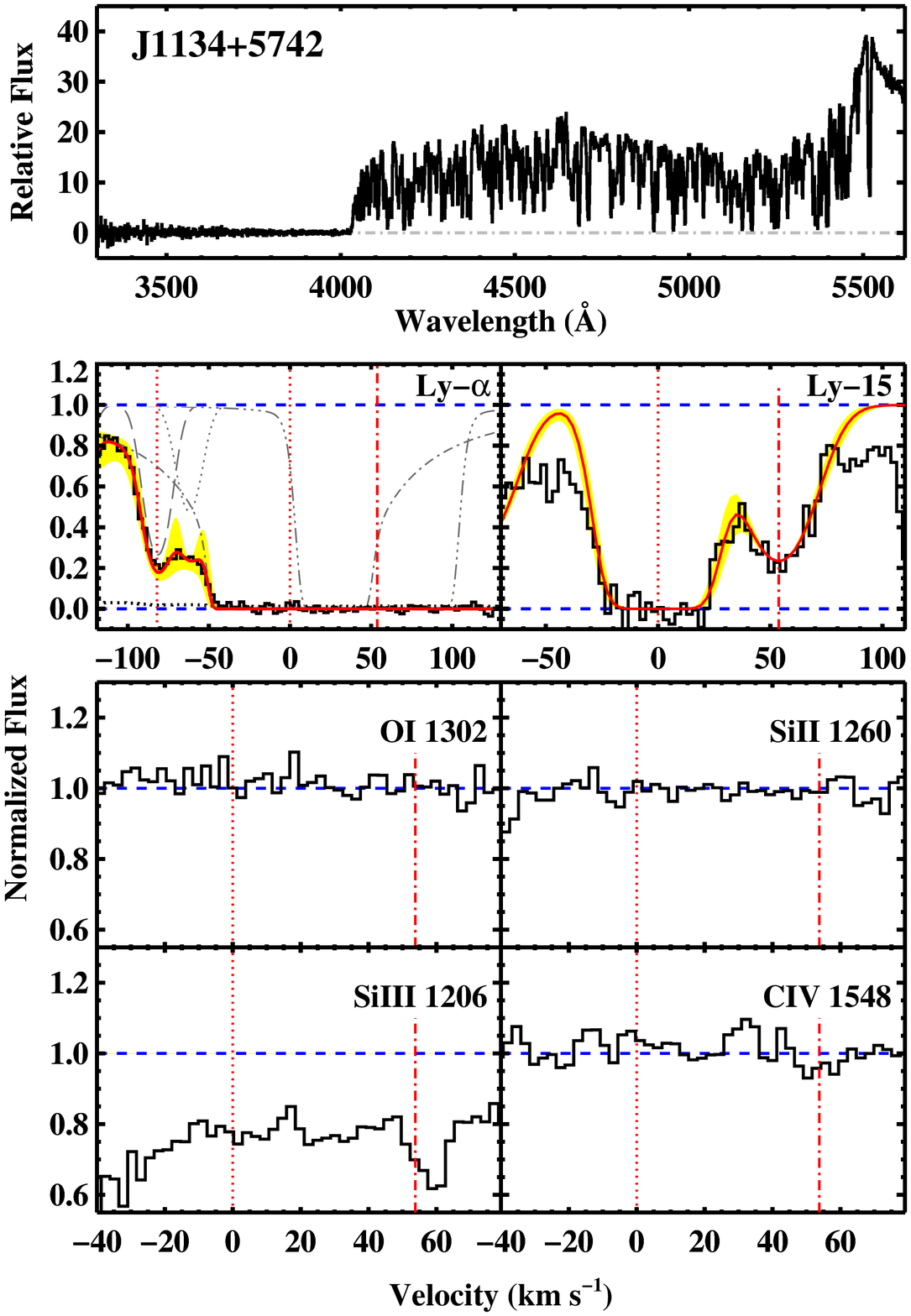}
\end{center}
\end{figure*}
\noindent
{\bf Figure \figllso:} 
{\small Top panel: Keck/LRIS spectrum of the QSO \nameone. A LLS at $z\sim3.4$ is 
clearly visible from the break at $\sim 4000$ \AA. 
Middle panels: \ion{H}{I} Lyman series transitions in the \llso. Superimposed to the
data are the best-fit model (red line) and the 2$\sigma$ errors (yellow shaded regions).
Individual components included in the model are marked with thin gray lines 
and the position of the hydrogen and deuterium are indicated by vertical dotted lines. 
The second hydrogen component (LLS1134b) at $+54~\rm km~s^{-1}$
is marked with a dash-dotted line. Bottom panels: selected strong metal-line 
transitions in  \llso\  (dotted lines) and LLS1134b (dash-dotted lines). 
Unrelated absorption from the IGM contaminates the \ion{Si}{III} 
transition (see Fig. \figmetline).}

\clearpage
\begin{figure*}[!ht]
\begin{center}
\includegraphics[scale=0.6]{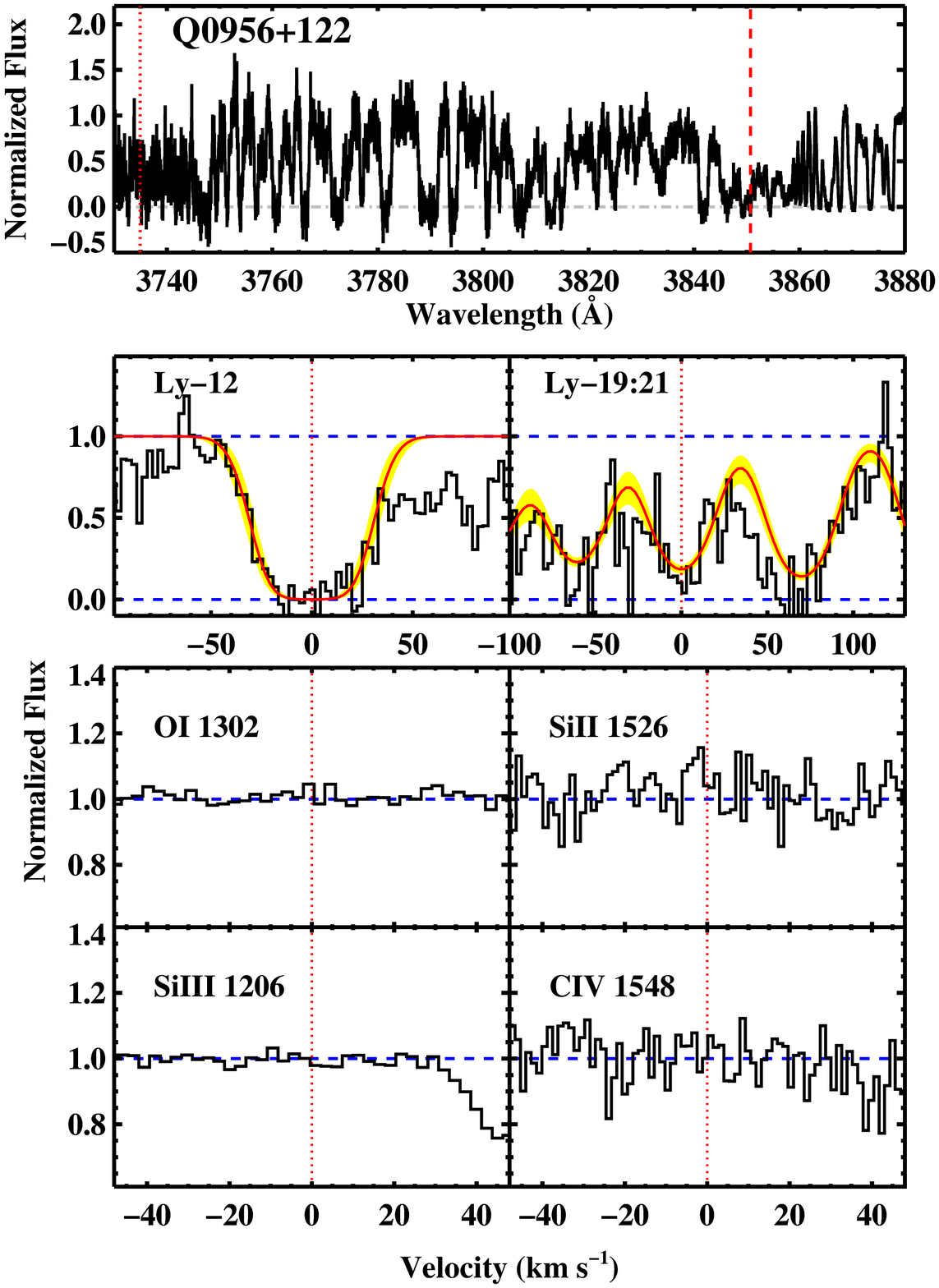}
\end{center}
\end{figure*}
\noindent
{\bf Figure \figllsta:}
{\small Top panel: Keck/HIRES spectrum of the QSO \nametwo. The flux 
decrements at $\sim 3860$ \AA\ and $\sim 3750$ \AA\ reveal 
two partial LLSs at $z\sim3.22$ (\llsta) and  $z\sim3.10$ (\llstb). 
The corresponding Lyman limits are marked
with vertical lines. Middle panels: hydrogen Lyman series transitions
for \llstb\ shown relative to \zllstwo. 
Superimposed to the data, the best-fit model (red lines) and $2\sigma$ uncertainties
(yellow shaded regions). 
Bottom panels: selected strong metal-line transitions for \llstb.}

\clearpage
\begin{figure*}[!ht]
\begin{center}
\includegraphics[scale=0.6,angle=90]{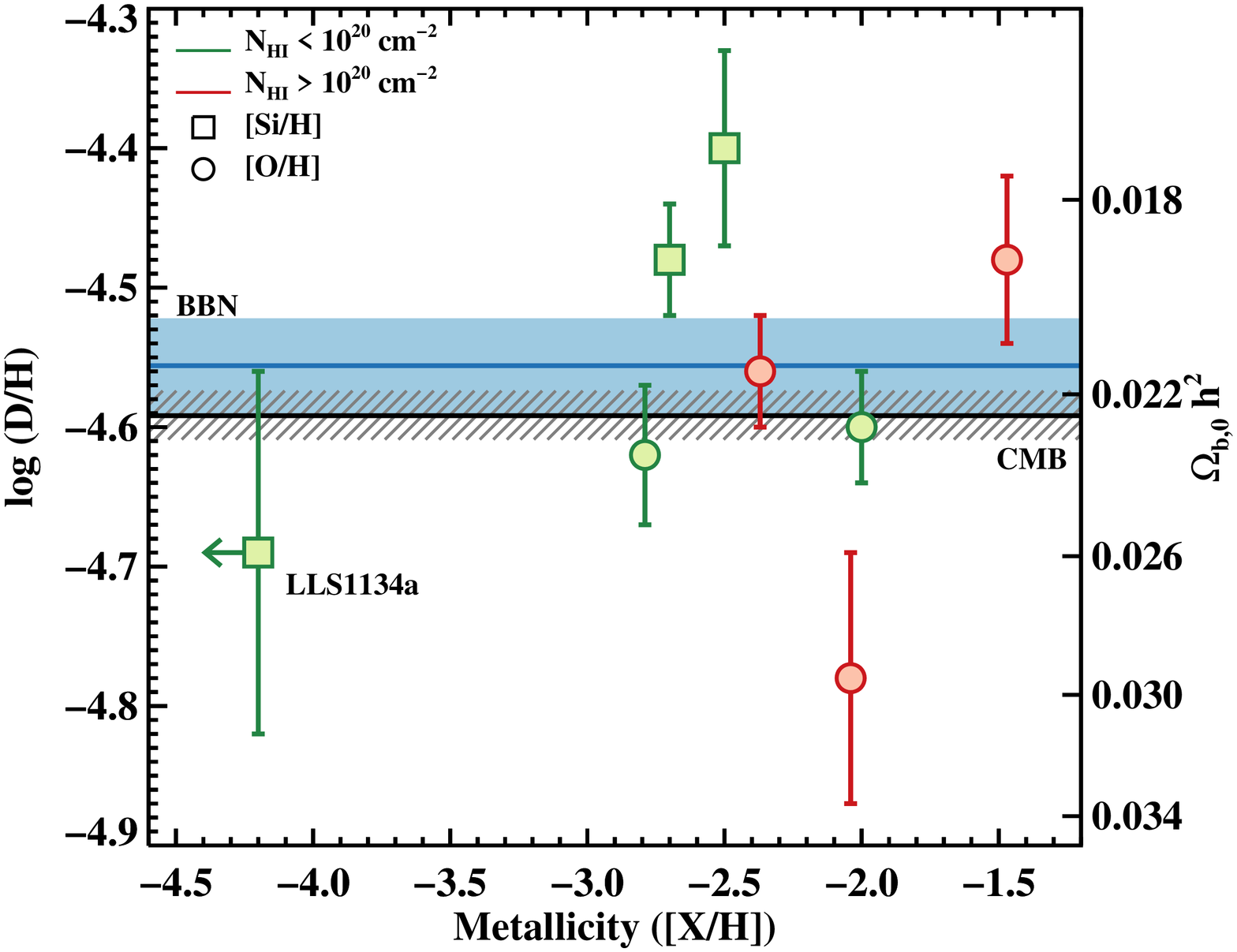}
\end{center}
\end{figure*}
\noindent
{\bf Figure \figdh:}
{\small Deuterium abundances as a function of metallicity 
($\rm [X/H] \equiv \log (X/H) - \log (X/H)_\odot$) for the $z>2$ absorption line systems.
Green symbols are for LLSs, while red symbols are for higher column density 
absorbers ($N_{\rm HI}> 10^{20}~\rm cm^{-2}$). Metallicities obtained with silicon 
are indicated by squares, while those obtained with oxygen are indicated with 
circles. The right-hand axis translates the deuterium abundance to the 
cosmic baryon density $\Omega_{b,0} h^{2}$. The inferred $\Omega_{b,0}h^{2}(\rm BBN)$  is
shown with a solid blue line, together with the 1$\sigma$ errors (light blue shaded area).
The CMB value and 1$\sigma$ errors from WMAP7 are instead shown with a solid black line  
and a gray dashed area.}

\clearpage
\begin{figure*}[!ht]
\begin{center}
\includegraphics[scale=0.58,angle=90]{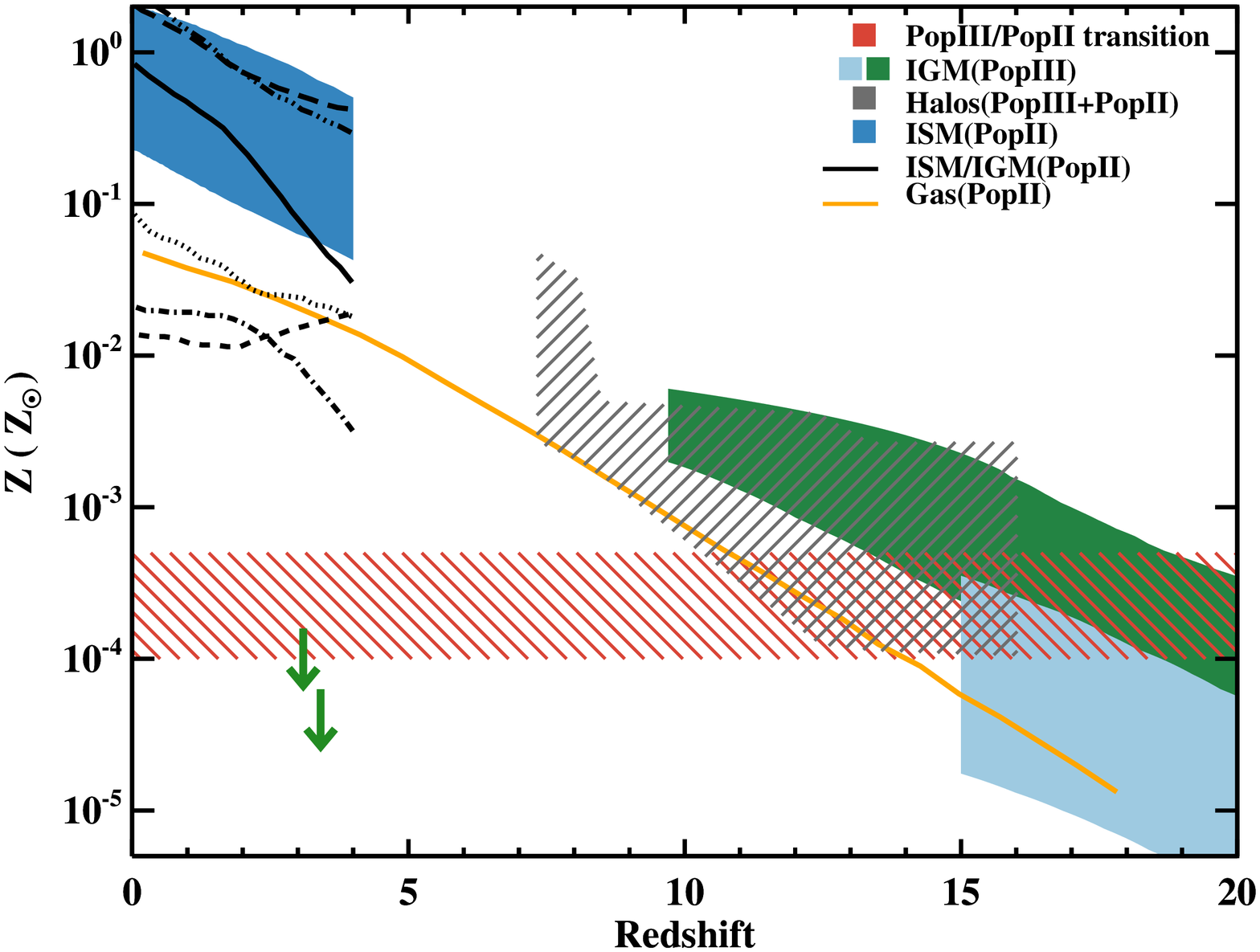}
\end{center}
\end{figure*}
\noindent
{\bf Figure \figpreenr:} 
{\small Overview of analytic models and simulations for the metal enrichment of the universe. 
Light blue and green shaded regions: IGM metallicity from PopIII 
stars with mixing between $1-0.05$ and different star formation histories \cite{mac03,yos04}.
Red dashed region: critical metallicity that marks the transition 
between PopIII and PopII stars \cite{bro01,sch02}. Orange line:
analytic model for the gas metal content in the universe from PopII stars 
and galactic winds \cite{her03}. Gray shaded region: gas metallicity
within halos from hydrodynamical simulations that include yields from both PopIII 
and PopII stars \cite{wis10}. 
Blue shaded region: analytic model for the ISM metallicity 
at different halo masses ($10^{11}-10^{14} \rm M_{\odot}$) and different 
wind models \cite{dav11}. Black lines: metallicity from hydrodynamical simulations 
with momentum driven winds \cite{opp11} in condensed gas (solid line),
hot halo (dotted line), warm-hot intergalactic medium (dashed line), 
diffuse gas (dash-dotted line), ISM (dash-triple-dotted line), and
stars (long-dashed line). Upper limits on the metallicities of \llso\ and \llstb\ are marked 
with green arrows.}

\clearpage
\setcounter{page}{1}

%
%
%
%
%
%
%

\noindent {\Huge \bf Supporting Online Material} 

\section{Observations and Data Reduction}\label{sec:obs}

Previous spectra of quasars \nonelong\ and \nametwo\ had shown
substantial absorption at wavelengths $\lambda < 4000$\AA\ 
characteristic of the continuum opacity of \ion{H}{I} gas.  
\llso\ and \llstb\ were identified as part of an ongoing 
program to explore the incidence and physical 
characteristics of LLSs \cite{prt10,ome07}
from a sample of $\sim 50$ absorbers,
many of which have apparently low metallicity ($<1/100$ solar).

We first observed quasar \nameone\ on UT January 5, 2006
with the HIRES spectrometer \cite{vog94} on the Keck\,I 
telescope on Mauna Kea. The data were obtained in two exposures totaling 
6300 seconds, with the instrument configured with the red
cross-disperser, the C5 decker, and the kv380 blocking filter to
suppress second order flux.  In this configuration, the instrument
provides a resolution of $\sim 8$ \kms\ FWHM, and the echelle and
cross-disperser angles were set to provide wavelength coverage 
$4017$ \AA\ $< \lambda < 8534$ \AA.  These spectra have gaps in
coverage at $\lambda
\sim 5450$\AA\ and 7050\AA\ due to spacings between the CCDs in the
detector mosaic.  There are also a series of gaps in wavelength
coverage beyond $\lambda
> 6300$\AA\ where the free spectral range of the spectrometer exceeds
the detector width.

We observed \nametwo\ with HIRES once on UT January 3, 2006 for 7200 
seconds, and again on UT April 7, 2006 for 1800 seconds.  
For the January data, HIRES was configured with the
UV cross-disperser and the C5 decker, again providing $\sim 8$ \kms\
FWHM resolution.  The wavelengths covered in this
configuration were $3621$ \AA\ $< \lambda < 5533$ \AA, with gaps in
coverage only related to the CCD mosaic.
The April 2006 data had HIRES configured with the red
cross-disperse, the C1 Decker, and the og530 blocking filter.  The
wavelengths $5465$ \AA\ $< \lambda < 9998$ \AA\ were covered with wavelength
gaps due to the spacings of the CCD mosaic and the limited detector
size.

For both quasars, the 2D images were reduced with the HIRESredux
pipeline\footnote{http://www.ucolick.org/$\sim$xavier/HIRedux/index.html}
which extracts and coadds the data.  An interactive continuum fitting
procedure within the pipeline assigns a continuum level to each
coadded spectral order, and the continuum normalized spectra are then
combined into a single 1D spectrum.  For \nameone, the resultant
1D spectrum has a signal-to-noise ratio S/N~$\sim 25$ per 2.6 \kms\ pixel at
6300\AA, the central wavelength.  For \nametwo, the 1D spectra have S/N~$\sim 30$ 
per 2.6 \kms\ pixel at 4500\AA\ for the UV cross-disperser data, and $\sim 10$ 
per 1.4 \kms\ pixel at 6000 \AA\ for the red cross-disperser data.

We also observed \nameone\ on UT 2011 July 1 with the LRIS
spectrograph \cite{oke95} on the Keck\,I telescope on Mauna Kea.
We obtained a single exposure of 400 seconds, with the instrument configured with 
the 600/4000 grism, the D560 dichroic and the 600/7500 grating, tilted to ensure 
continuous spectral coverage between the blue and red arms. 
The 2D images were reduced with the LowRedux 
pipeline\footnote{http://www.ucolick.org/$\sim$xavier/LowRedux/index.html}
which calibrates and extracts the data.

\section{Hydrogen and Deuterium Analysis}

We now describe the \ion{H}{I} and \ion{D}{I} analysis for each LLS.  

\subsection{The Lyman Limit System at \zllsone\ toward \nameone}\label{sec:LLSone}

Both the strong Lyman limit absorption 
visible in the Keck/LRIS spectrum (Fig. \figllso), 
and the saturation of the high-order Lyman series lines 
in the Keck/HIRES data (Fig. \figlymano)
imply a large \ion{H}{I} column density,
\nhi. Fig. \figlris\ shows a zoom-in of the Keck/LRIS
spectrum for \nameone, focusing on the Lyman limit absorption.
We estimate the flux just redward of the Lyman limit to be $13 \pm 2$
in our relative units.  This estimate appropriately includes the attenuation of the
quasar light by the \lya\ forest.  In the following, we assume a
flat spectrum for the continuum and note that a more realistic 
spectrum would yield a slightly higher estimate for the \nhi\ value.  
Just blueward of
the Lyman limit ($\lambda = 3950-4000$\AA),  we set a conservative
upper limit to the relative flux of 0.05.  Combining these two
measurements, we estimate a lower limit to the optical
depth at the Lyman limit of $\tau_{\rm LL} > \ln(9./0.05) > 5$ for a 
total $\mnhi > 10^{17.9}
\cm{-2}$.  The model for this \nhi\ limit, shown as the green line in
Fig. \figlris, is marginally acceptable at $\lambda \sim
4000$\AA, but predicts a recovery of the flux at $\lambda < 3800$\AA\
that appears inconsistent with the observations.
Metal-lines and strong \lya\ lines in the HIRES spectrum, however, reveal 
the presence of other absorbers along the sightline to \nameone\ at redshifts 
$z=3.0283$, $z=3.0753$, $z=3.1047$, $z=3.2396$, $z=3.2616$, and $z=3.3223$. 
These systems could contribute to the opacity at $\lambda < 3942$\AA.
If they do not, then \llso\ must have a higher \nhi\ value.

Fig. \figlris\ shows the data binned with a weighted mean 
in windows of 50\AA\ between $3400-4000$\AA.
Overplotted on these binned evaluations (whose errors only
reflect statistical uncertainty) is a model curve for
$\mnhi = 10^{18.05} \cm{-2}$ (blue line).  This \nhi\ value reproduces the
observed absorption to $\sim 3400$\AA.
Larger \nhi\ values are not ruled out by the
analysis, especially when one allows for systematic error (e.g.\ sky
subtraction). 
In summary, we report a lower limit to the total \nhi\ value of
$10^{17.9} \cm{-3}$ and note the data require yet higher values if
\llso\ dominates the opacity at $\lambda < 3700$\AA.

To further refine the \nhi\ measurement and its redshift distribution,
we have performed detailed analysis of the Lyman series lines.  This
analysis, however, is challenged by the fact that gas with $\mnhi \sim
10^{18} \cm{-2}$ has Lyman series lines that lie on the saturated
portion of the curve-of-growth (e.g.\ \cite{prt10}).  
As such, we proceed cautiously as this may impose large uncertainties in the model.
We may first set a strict upper limit to the total \nhi\ of \llso\ from the
absence of substantial damping wings in the \lya\ profile (Fig. \figllso).
Because the flux would be over-absorbed for all models with 
\nhi\ exceeding $10^{18.7} \cm{-2}$,
this is the largest \nhi\ value allowed for \llso.
Second, the complete absorption at $\lambda \sim 4030$\AA\ by the
Lyman series requires $\mnhi > 10^{17.6} \cm{-2}$, consistent with the
Lyman limit analysis performed above.  

Lastly, we can model the \ion{H}{I} 
absorption using the {\sc vpfit v9.5} package that allows one to 
fit multiple Voigt profiles to the spectrum. 
Our general approach is to include as few components 
as possible (each defined by a redshift $z$, column density \nhi, 
and Doppler parameter $b_{\rm HI}$) 
to match the Lyman series absorption. 
The data are best described by two principal \ion{H}{I}
components (\llso\ and LLS1134b), as evident from a visual inspection of 
the higher order Lyman series lines (Fig. \figllso\ and \figlymano). 
Since  LLS1134b is unsaturated beyond
Ly-14, we first measure the line-parameters of this subsystem
using the Lyman series lines between Ly-10 and Ly-16.
We find  $z=3.41167\pm0.00001$, $b_{\rm HI}=18.0\pm0.9~\rm km~s^{-1}$ 
and \lnhi $=16.71\pm 0.02$
indicating that this component has a nearly negligible contribution to 
the observed Lyman limit. Adding an additional  uncertainty of $0.02$ dex
related to the $1-3\%$ error in the continuum placement, 
our best fit value for the column density of LLS1134b becomes \lnhi $=16.71\pm 0.03$.
Then, we model \llso\ and the deuterium 
absorption, using both high order Lyman series lines (Ly-7 through Ly-16) 
as well as Ly-$\alpha$ and Ly-$\beta$.  For these transitions,
we must include absorption from gas unrelated to LLS1134 which we ascribe to
coincident \lya\ absorption from lower redshift absorbers.
A summary of the model parameters is given in Table \tabfitllso.

The strongest \ion{H}{I} component has $z=3.410883\pm0.000004$,
\lnhi $=17.94\pm 0.05$ and an \ion{H}{I} Doppler parameter 
$b_{\rm HI}= 15.4\pm0.3~\rm km~s^{-1}$. 
The latter is consistent with a predominantly ionized gas 
at a temperature $T \le 1.4 \sci{4}$\,K.
The line-parameters for \llso\ are well
constrained by the full Lyman series and the resultant \nhi\ value is
consistent with our Lyman limit analysis of the Keck/LRIS spectrum.  
Despite the line-saturation,  there is substantial constraint on the
$b_{\rm HI}$ value, and thereby the \nhi\ value, from the line-profile shapes
of the Lyman series lines.  This conclusion hinges, however, on the
assumption that the \ion{H}{I} absorption at $z\sim 3.41088$
is dominated by a single component.   As we will demonstrate in the next paragraph, 
this assumption is well supported by the analysis of neighboring \ion{D}{I}
absorption. From the above discussion, we conclude $\mnhi = 10^{17.95 \pm 0.05} \cm{-2}$ 
for \llso. The $0.02$ dex error on the continuum placement is in this case 
negligible compared to the statistical uncertainty.
In regards to the metallicity of this gas, 
we further emphasize that \lnhi $\ge 17.95$ is preferred by the 
analysis of the Lyman continuum opacity, despite the symmetric 
statistical error. If anything, metallicity limits derived from this
\nhi\ value may be considered conservative. 

In addition to the \ion{H}{I} Lyman series, 
\llso\ exhibits substantial absorption for the \ion{D}{I} \lya\ and
\lyb\ transitions at the expected $-82~\mkms$ offset from the
\ion{H}{I} lines. Due to blending with absorption lines from lower 
redshift gas, deuterium is not detected in the remaining Lyman series lines. 
Although both of the \ion{D}{I} transitions are partially blended with (presumed) 
\ion{H}{I} absorption, a simultaneous fit to the 
hydrogen and deuterium absorption lines (Fig. \figlymano)
constrains $\log N_{\rm DI} = 13.26 \pm 0.04$ and $b_{\rm DI}=10.2\pm0.8$.
The latter parameter is fully consistent with the $b_{\rm HI}$ value
derived from the hydrogen absorption. The deuterium estimate is 
more sensitive to the continuum placement error than hydrogen, 
particularly at the position of Ly-$\beta$. Therefore, we include an additional 
uncertainty of $\sim 0.08$ dex, computed 
by repeating multiple times this analysis after having varied the 
continuum level by $\pm 1-3\%$.
When we express the Doppler parameter as a 
function of thermal broadening $b_{\rm ther}$ and turbulent component $b_{\rm turb}$, 
the expected ratio of the hydrogen and deuterium $b$ parameters is 
$b_{\rm DI}/b_{\rm HI}=\sqrt{1-0.5 \phi}$ with  
$\phi=b_{\rm HI, ther}^2/(b_{\rm HI,ther}^2+b_{\rm turb}^2)$ and $0<\phi<1$. 
The measured $b_{\rm DI}/b_{\rm HI}=0.66$ is within 10\% of the expected value for 
turbulent broadening ($\phi \sim 1$).
The best-fit $\rm D/H$ ratio for \llso\ is $\log \rm (D/H) = -4.69 \pm 0.06$,
in agreement with previous determinations (Table \tabdeutsys).
This consistency provides additional confidence on the model for the 
hydrogen absorption. Due to the partial blending of the \ion{D}{I} lines with intervening 
hydrogen, the deuterium column density is affected by the 
additional uncertainty on the line parameters for the hydrogen absorption. 
Including this error, our best estimate becomes $\log \rm (D/H) = -4.69 \pm 0.09$.
Finally, including the uncertainty on the continuum determination, 
we find $\log \rm (D/H) = -4.69 \pm 0.13$.

\subsection{The Lyman Limit System at \zllstwo\ toward \nametwo}\label{sec:LLStwo}

A partial LLS (\llsta)  was previously identified in low-resolution
spectra of  \nametwo\  and associated with corresponding metal-line absorption at 
$z=3.2228$ \cite{ste90}.  We confirm this identification (Fig. \figllsta).
An LLS at lower redshift was also reported by Steidel et al. \cite{ste90}, 
who proposed it was associated with a metal-line system at 
$z=3.1142$.  Our analysis of the HIRES spectra, 
however, only reveals substantial high-order Lyman series absorption for a
system at $z \sim 3.096$ and we adopt this as the redshift for \llstb.

Non-zero flux is detected blueward of the Lyman limit for each system, 
implying $\mnhi < 10^{17.8} \cm{-2}$ for both \llsta\ and \llstb.
At this optical depth ($\tau_{\rm LL}\sim 1$), the total \nhi\ value is well established by
the Lyman limit opacity and the analysis of unsaturated Lyman series
lines.

The hydrogen Lyman series of \llsta\ is detected through to Lyman-21
in our HIRES data, together with \ion{C}{IV} and \ion{Si}{IV} metal
lines.  By modeling the unsaturated hydrogen transitions from Lyman-16
to Lyman-21, we derive the redshift $z=3.223194\pm0.000002$,
the column density \lnhi $=17.37\pm0.01$, and the Doppler parameter
$b_{\rm HI}=20.4 \pm 0.2$. Similarly, for \llstb, the hydrogen Lyman
series is clearly visible through to Lyman-15 and, at lower
signal-to-noise, through to Lyman-21. No other metal lines are
detected at the hydrogen position, while deuterium is blended in the
strongest transitions. Combining both high-order saturated and
non-saturated \ion{H}{I} absorption lines, we model this LLS with a
single component at $z=3.096221\pm0.000009$, with 
$b_{\rm HI}=20.2\pm0.8 ~\rm km~s^{-1}$ and \lnhi $=17.18\pm0.03$ (Fig.
\figllsta\ and \figlymant). After accounting for an additional $0.02$
dex of uncertainty on the continuum placement, 
the best fit column density for \llstb\ becomes \lnhi $=17.18\pm0.04$.
The Doppler parameter implies a temperature $T
\le 2.5 \times 10^{4}$ K, typical for photoionized gas. 

\section{Metal Line Analysis}

We now discuss upper limits to the column densities of atoms and ions
for heavy elements in these metal free LLSs.

\subsection{LLS1134}

\llso\ is conspicuously free of metal-line absorption, as shown in
Fig. \figllso\
where we mark for selected ionic transitions the position corresponding 
to this LLS and to the weaker hydrogen component 
(\zllsone\ and $z=3.41167$ respectively). For both \llso\ and 
LLS1134b, we can place upper limits to the ionic column
densities using the apparent optical depth method \cite{sav91}, 
with the key limiting factors being the $\rm S/N$ of the spectrum 
and contamination from the \lya\ forest (i.e.\ \lya\ lines at
unrelated wavelengths). For this analysis, we 
choose a velocity window of $\pm 15 ~ \rm km~s^{-1}$ that is 
wide enough to encompass the expected width for metal lines
given the hydrogen and deuterium absorption properties.
A larger velocity window of $\pm 20 ~ \rm km~s^{-1}$ would 
yield slightly higher ($\sim 0.06$ dex) limits.
A summary of the metal column densities is presented in Table \tabmetlim.

For most of the listed ionic transitions, we place robust upper limits 
based solely on the variance and the 
uncertainty associated with a $1-3\%$ ($\rm S/N$ dependent) error
in continuum placement. 
Transitions blueward of 5450\AA, however, lie within the
\lya\ forest which complicates the estimates. 
Of particular interest to the metallicity of \llso\ are the strong
transitions of \ion{Si}{III}~1206 and \ion{C}{III}~977.  Fig. \figmetline\ 
reveals that there is substantial absorption at the predicted
wavelength for each transition.  This absorption is broad, however,
and has peak optical depth that is substantially offset from the
predicted line-center.  It is certain, therefore, that these
transitions are blended with coincident absorption by the IGM.  In
this case, an upper limit based on the apparent optical
depth would overestimate the column densities.

Better limits to the \ion{Si}{III} and \ion{C}{III} column
densities can be derived using models for these absorption lines.
We fix the Doppler parameters for the metal lines using the 
hydrogen and deuterium $b$ values. From the \ion{H}{I} and \ion{D}{I} lines, we infer
a temperature $(1.6 \pm 0.2) \times 10^4$ K and an upper limit on the turbulent
velocity $b_{\rm turb}<2.2~\rm km~s^{-1}$. Carbon and silicon absorption lines 
that arise in gas with these properties would be characterized by $b_{\rm C} = 6.9~\mkms$
and $b_{\rm Si} = 5.3~\mkms$. These Doppler parameters are within the range 
commonly observed for metal-line transitions (e.g.\ \cite{lid10}).
A first limit to the column densities of  \ion{C}{III} and \ion{Si}{III}
is set by demanding that the peak optical depth of each metal line does not
exceed the observed absorption. This approach,
that basically ignores the fact that the opacity is
dominated by coincident IGM lines,  yields upper limits
of $\log N_{\rm SiIII} < 11.85$  and $\log N_{\rm CIII} < 12.45$.
Lines of column density higher than these would exceed the observed
absorption at the corresponding transitions, as shown in the top panels of 
Fig. \figmetline. 

A substantial fraction of the opacity at the \ion{Si}{III} and \ion{C}{III}
frequencies must however arise from the two broad  Ly$\alpha$ forest lines
blueward to these metal transitions.  
We set therefore limits to $\log N_{\rm SiIII}$ and $\log N_{\rm CIII}$ using a two 
component model that includes both the metal lines and the Ly$\alpha$ forest. 
Following this procedure, we find $\log N_{\rm SiIII}<11.40$ and 
$\log N_{\rm CIII}<12.20$. Since additional 
Ly$\alpha$ lines at both negative and positive velocities 
are required to account for all the observed absorption,  
the choice of this minimal scenario (only two components) yields a 
quite conservative limit to the metal column densities.

Inspecting the strongest transitions of LLS1134b, the second main hydrogen component 
at $z=3.41167$, we see that \ion{Si}{III} and \ion{C}{III} 
are detected. These transitions are commonly found in highly 
ionized LLSs. Again using the apparent optical depth method, we
find for the column densities of these ions $\log N_{\rm CIII} =
13.03 \pm 0.02$ and $\log N_{\rm SiIII} = 12.15 \pm 0.02$.
While a strong line is clearly visible at the position of \ion{S}{III}
for this component, the implied column density would greatly exceed that for
\ion{Si}{III} and \ion{C}{III} indicating that the absorption is dominated
by contamination from the \lya\ forest.

\subsection{\llstb}

Similarly to \llso, \llstb\ is free of metal line absorption
(Fig. \figllsta). Using the apparent optical depth method, we determine upper 
limits to the ionic column densities in the velocity interval 
$\pm 15 ~ \rm km~s^{-1}$, consistent with the expected metal line width.
The metal column densities for the strongest transitions, 
including the variance and uncertainty related to continuum placement,  are listed in 
Table~\tabmetlim.

\ion{C}{III} $\lambda 977$ is blended in the \lya\ forest. 
Supported by the lack of any silicon absorption (\ion{Si}{II}, \ion{Si}{III}, 
and \ion{Si}{IV}) and the absence of \ion{C}{II} and \ion{C}{IV}, we conclude 
that carbon is not present in this system to limits comparable to the ones 
inferred from silicon. Therefore, we do not include \ion{C}{III} $\lambda 977$
in our estimates of the metallicity, but we rely on the \ion{Si}{III} upper limit
which we measure directly with the apparent optical depth method.
We note that a larger velocity window of $\pm 20 ~ \rm km~s^{-1}$
would result in a slightly higher ($0.06$ dex) column density limit. 
We also note that the strongest \ion{Si}{II} line 
at $\lambda 1260$ is overwhelmingly contaminated by intervening
\ion{H}{I} absorption and we use the \ion{Si}{II} $\lambda 1526$ 
line to set a limit for this ion.  

\section{Metallicity Limits}

Having established the \ion{H}{I} column densities and
limits to the ionic column densities of Si, C, and O, we may set
upper limits to the heavy element abundances in the two LLSs.
This requires, however, a careful consideration of
the ionization state of the gas.  
Generally, the ionization state is assessed through a comparison of
observed ionic column densities with ionization models.  Absent the
detection of {\it any} metal transitions, we must consider
physically-motivated scenarios for the properties of the gas and
thereby estimate limits to the metallicities.

In a perfectly neutral medium, all of the elements would be in their atomic
state (ignoring molecular formation) and one could estimate the
metallicity directly from the observed column densities of the atomic
transitions (e.g.\ \ion{H}{I}, \ion{C}{I}, \ion{O}{I}).
In this extreme model one would recover 
$\rm \log(O/H) - \log(O/H)_\odot \equiv [O/H]~<-1.9 (-1.4)$ 
and $\rm [C/H]~< -2.1 (-1.3)$ 
for \llso\ (\llstb), which are
the most conservative limits possible.
In essentially all astrophysical environments, however,
the gas is irradiated by local and external sources 
which photoionizes at least the outer layers of the cloud.
Photons with energy $h\nu < 1$\,Ryd will photoionize all
elements whose first ionization potential is below that of hydrogen
(e.g.\ C, Si) and harder photons may ionized a majority of the
\ion{H}{I} gas and place the heavy elements in higher ionization
states. Collisional processes, predominantly with electrons and hydrogen nuclei, 
may further ionize the medium for gas at $T> 10^4$\,K.  

Focusing on photoionization first, the population of galaxies
and quasars at $z \sim 3$ is known to generate an extragalactic UV
background (EUVB) which ionizes the majority of baryons in the
universe (e.g.\ \cite{haa96,rau97}).  For gas with the relatively
low \ion{H}{I} column densities of our LLSs, one predicts that the
EUVB has substantially ionized the hydrogen gas and any heavy
elements that are present.
To quantitatively assess this process and its impact on the
metallicity estimates, we have calculated a series of photoionization
models using the {\sc cloudy} software package \cite{fer98}.  We assumed
that each LLS may be modeled as a series of layers with constant
hydrogen density $n_{\rm H}$, constant and low metallicity, 
and a plane-parallel geometry.  We
irradiated this gas with the Haardt and Madau EUVB model\footnote{The results would be
slightly different with a different shape for the radiation field,
but the differences are much smaller than those related to varying
the ionization parameter.} (quasars+galaxies)  \cite{haa11}
evaluated at the redshift of each LLS 
(e.g.\ $J_\nu = 2.45 \sci{-22} \; {\rm erg \, s^{-1} \, Hz^{-1} \,
  cm^{-2}}$ at 1\,Ryd\ for $z=3.4$)
and considered gas with a range of \nh\ values.  
It is standard practice to
describe a given photoionization model by the ionization parameter 
$U \equiv \Phi/(\mnh c)$ with $\Phi$ the flux of ionizing photons.
Each model was restricted to have a total \nhi\ column density
consistent with the observations.

In Fig. \figcla, we present the upper limits
to the metallicity of the two LLSs as a function of \nh\ values.
The figure indicates the limit imposed for the range of
ions constrained by our observations.  
As a function of \nh,
each curve shows the metallicity limit from a given
ion according to the measured upper limit on its column density.  The black
symbols then trace the lowest metallicity imposed by the full set of
measurements at each \nh\ value.  In all cases we compare to the solar
abundances of \cite{asp09}.

At high densities (low $U$
values), the tightest limits are given by the lowest ionization stages of
the heavy elements (Si$^+$, O$^0$, C$^+$).  
These set upper limits to the metallicity of $\lesssim 10^{-3} \zsun$.
We emphasize, however, that 
such high densities are improbable. 
A gas with $\mnh > 10^{-1} \cm{-3}$ is predicted to have a
neutral fraction $\mxhi \equiv \mvhi/\mnh > 0.1$,
but all of the LLSs that have been studied to date exhibit
substantial absorption from higher ionization states of Si, C, and O.
When analyzed as photoionized gas, one derives ionization parameters
$\log U = -3$ to $-1$ (Fig. \figudis).
At the respective redshift for the two LLSs, this implies a volume density 
$\mnh < 10^{-2} \cm{-3}$ which corresponds to an overdensity relative to the mean 
baryon density of $\rho/\rho_{\rm mean} < 850$.  
As \nh\ decreases (and $U$ increases), the heavy 
elements shift to higher ionization states and the limits on the gas
metallicity become much more stringent.  This is partly a result of
tighter limits to the ionic column densities but it is primarily
because our models are forced to match the observed \nhi\ value.  
At higher ionization
parameter, the gas is more highly ionized and the total implied
hydrogen column density is correspondingly larger ($N_{\rm H} = \mnhi /
\mxhi$).  Therefore, at fixed metallicity this implies higher total 
column densities for the heavy elements. 
For the ionization
parameters typically measured for LLSs ($U > 10^{-3}$), we set an
upper limit to the gas metallicity (based on Si) of $Z < 10^{-4.2} (10^{-3.8}) \mzsun$
for \llso\ (\llstb).
This is a conservative value and also the most physically-motivated limit
to the metallicity.

If the intensity of the ionization field is much higher than the 
assumed EUVB, the lack of substantial \ion{Si}{IV} and \ion{C}{IV}
absorption imposes a tight limit to the metallicity, as evident
from Fig. \figcla. 
Extreme scenarios in which most of the metals are in even higher 
ionization states are highly implausible since at $\log U \gg 1$ 
a neutral fraction $x_{\rm HI} \ll -6$ would imply a total 
hydrogen column density $N_{\rm H} \gg 10^{24}~\rm cm^{-2}$. 
Further, if these two LLSs lie in proximity to 
a star forming galaxy, the ionization field is dominated by a
softer spectrum. To explore this possibility, we have calculated a
second set of \textsc{cloudy} models adding to the EUVB the
contribution from a galaxy with star formation rate 
$100~\rm M_\odot ~yr^{-1}$. We have further assumed that the escape
fraction of ionizing radiation is $f_{\rm esc} = 0.1$ and that the 
LLSs are at 100 kpc from the star forming disks. Under these 
conditions, the ionization states of the metals do not differ substantially 
from our previous calculation. Indeed, at $N_{\rm HI} = 10^{18}~\rm cm^{-2}$, 
the inferred metallicity for \llso\ is $Z<10^{-4.0}$.
Therefore, we conclude that the assumed limits of $\sim 10^{-4}$
for \llso\ and \llstb\ are representative of the underlying gas metallicity 
in different plausible scenarios.

For gas with $T> 2\sci{4}$\,K, collisional ionization is an important
process for establishing the ionization state of the gas.  If one
assumes collisional ionization equilibrium, it is straightforward
to set limits on the gas as a function of the gas temperature
\cite{gna07}.  
For $T > 2.5 \sci{4}$\,K, the limits are yet lower than those suggested by the
photoionization modeling. At $T \sim 5 \sci{4}$\,K,
the lack of substantial \ion{Si}{III} and \ion{C}{III} constrains the 
metallicity at $Z < 10^{-6} (10^{-5}) \mzsun$ for \llso\ (\llstb), 
while at higher temperatures, $T \sim 10^{5}$\,K, 
the upper limits on the \ion{Si}{IV} and \ion{C}{IV} column densities 
correspond to 
$Z < 3 \times 10^{-7} (10^{-6}) \mzsun$ for \llso\ (\llstb).
At temperatures below $T \sim 2.5 \sci{4}$\,K,
we expect both collisional ionization and photoionization to contribute, with 
the latter dominating.

\section{Comparison with Other Observations}

To compare the limits on the metallicity for \llso\ and \llstb\ with other
observations, we search the literature for metallicity estimates of cosmic 
structures in the redshift interval $2\lesssim z \lesssim 5$.
The metal abundances in solar units as a function of the gas density are summarized 
in Fig. \figmetsum. Since we cannot guarantee completeness in our search, 
this figure  offers only a qualitative assessment on the interval of metallicity reported 
in one or multiple studies which include large samples. Here we provide a
description of the different data shown in the figure.

The range of densities and metals for the Ly$\alpha$ forest is taken from \cite{sim04},
who compile \ion{O}{VI} absorption statistics at $z\sim 2.5$ using 
the line-fitting technique. These metallicities are 
consistent with other studies that are
based on different analysis of the IGM (e.g.\ the pixel optical depth method 
\cite{agu08}). Upper limits and statistical fluctuations around or below $Z\sim 10^{-3}$ have 
been also reported at densities $\rho/\rho_{\rm mean} \sim 1$. These are represented in 
Fig. \figmetsum\ using a downward arrow.

For the LLSs (here loosely defined in the interval $\log N_{\rm HI}\sim 10^{16}-10^{20.3}~\cm{-2}$), 
we search in the literature for published metallicity and densities between  $z=1.5-4.5$ 
\cite{prt10,pro99,kir03,dod01,pro06,pro99a,ome01,bur98a,bur98b,kir00,cri04,lev03,rei93,sar90,lev03b}. 
The range plotted in Fig. \figmetsum\ brackets these observations. For this class of 
objects we also present the logarithmic mean and standard deviation based on our 
(nearly) complete compilation. In addition to the metallicity, we gathered 
information on the ionization parameters, used to produce Fig. \figudis.
ISM metallicity in damped Ly$\alpha$ systems has been 
measured with both quasars and $\gamma-$ray bursts (GRBs). 
In both cases, we adopt a number density $0.05-10\cm{-3}$ which is 
consistent with the densities for the ISM atomic phase
\cite{spi78,jorgenson10}.
For the quasar sample, we report the range of metallicity between $z=2-5$ 
from \cite{pro03}, extended at the lower end with the metal poor sample of 
\cite{pen10}. For the GRB sample we adopt instead the metallicities presented in \cite{pro07}. 

Spectroscopy of high redshift galaxies provide abundances from unresolved 
\ion{H}{II} regions in star forming galaxies. Here, we adopt the metallicity 
in two samples of $z\sim 2$ and $z \sim 3$ galaxies from \cite{erb06,man09}.
We adopt a range of densities typical of nearby \ion{H}{II} regions
\cite{ost89}. Several studies have revealed super-solar metallicity in the broad line regions
of quasars. Here, we adopt a compilation of metallicity from a large sample of $z>3.7$
quasars and  we choose a density interval typically measured in the broad line region 
\cite{ost89}.

Finally, we present the metallicity in Milky Way halo stars.
Stellar metallicity is often quoted using iron abundance, but 
the density of heavy elements is dominated by O, C, Ne, and other
$\alpha$-elements. We therefore consider  
the oxygen abundances presented in \cite{fab09}, extended at low 
metallicity with the star reported by \cite{caf11}. 
Although the inferred metallicity at the lower end 
suffers from considerable observational uncertainties \cite{fab09}, 
values at  $Z\sim 10^{-3}\zsun$ are common in large compilations \cite{sud11}
even for stars with extremely low iron abundances. In fact, only a single star has been reported 
at $Z\sim 10^{-4} \zsun$ \cite{caf11}.

\clearpage
\begin{figure*}[!ht]
\begin{center}
\includegraphics[scale=0.6,angle=90]{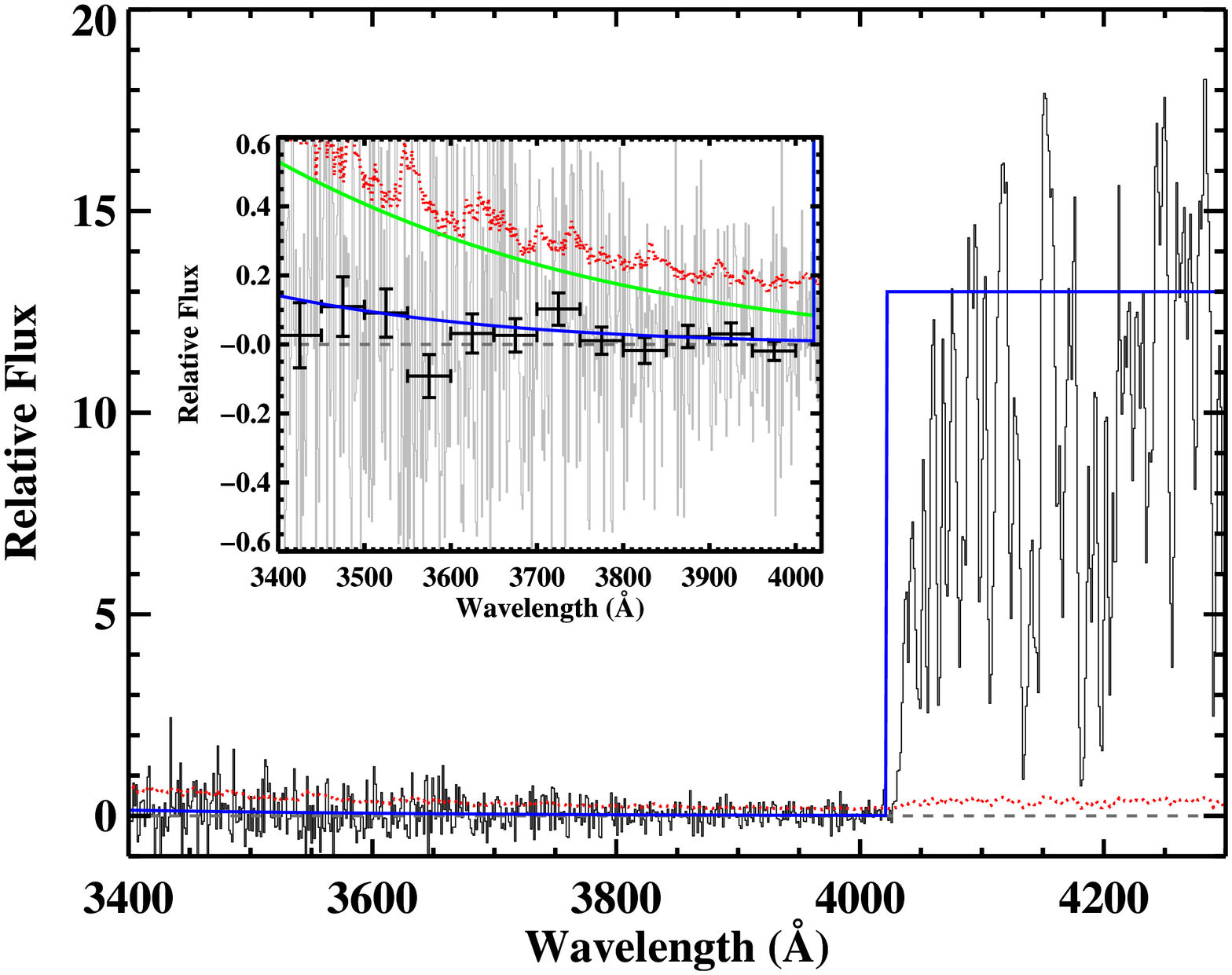}
\end{center}
\end{figure*}
{\bf Figure \figlris:} 
{\small  LRIS spectrum of \nameone\ at the Lyman limit. 
Superimposed to the data, we show the flux decrement 
for a $\mnhi = 10^{18.05} \cm{-2}$ LLS (blue line) together with the error on the flux
(red dotted line). In the inset, we show the data at full resolution 
(gray histogram) and in bins of 50 \AA\ (crosses). 
A model of the Lyman limit opacity 
for \lnhi$>17.90$ (green line) is also shown.}

\clearpage
\begin{figure*}[!ht]
\begin{center}
\includegraphics[scale=0.65]{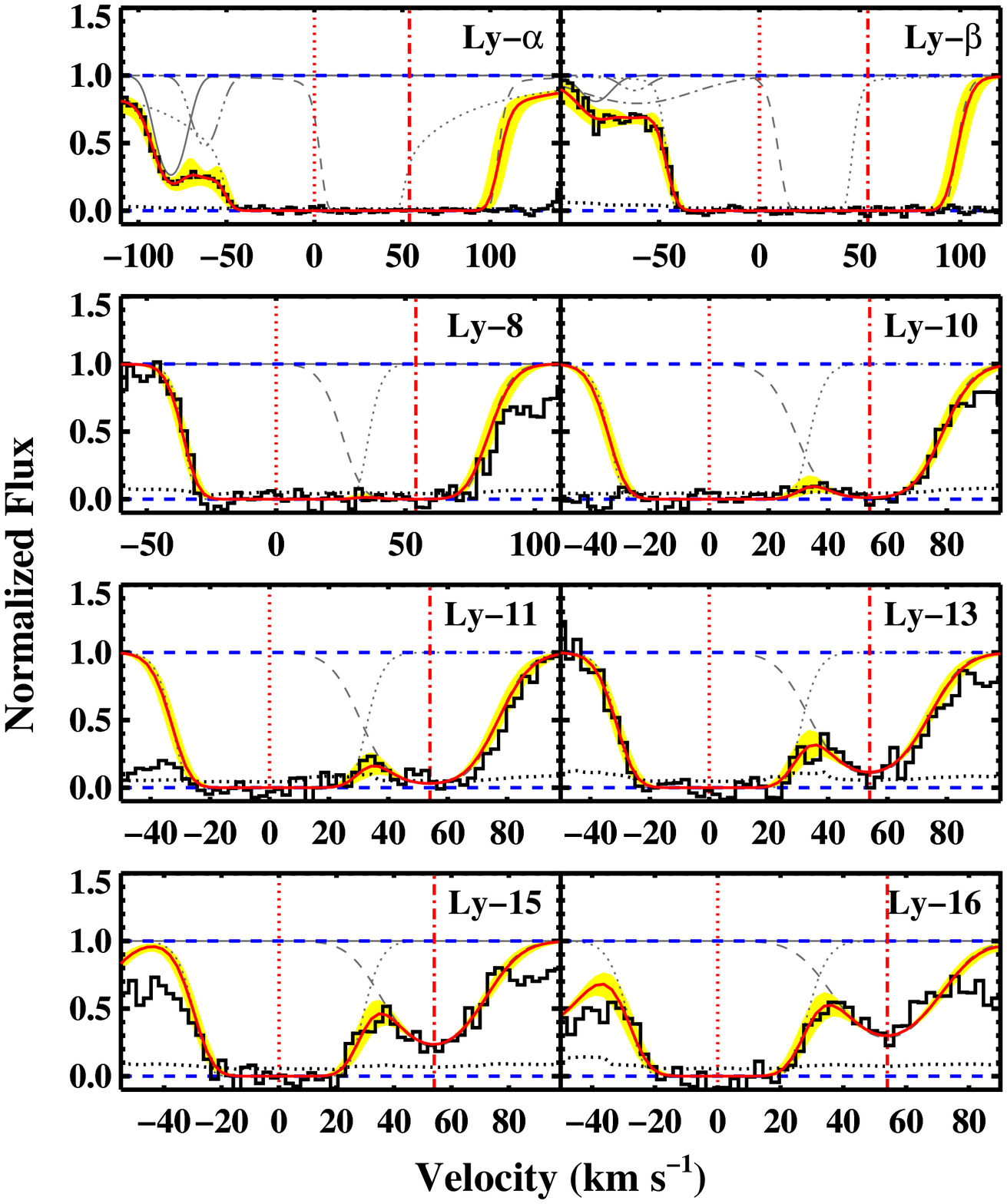}
\end{center}
\end{figure*}
{\bf Figure~\figlymano:}
{\small  Gallery of the Lyman series transitions (black histograms) 
and best fit models (red lines) for LLS1134. The yellow shaded regions represent the 
$2\sigma$ errors on the line parameters. 
\llso\ and LLS1134b are marked by vertical lines, while the dotted
black lines indicate the $1\sigma$ error on the flux. Individual components included in the
model are shown with gray lines.}

\clearpage
\begin{figure*}[!ht]
\begin{center}
\includegraphics[scale=0.65]{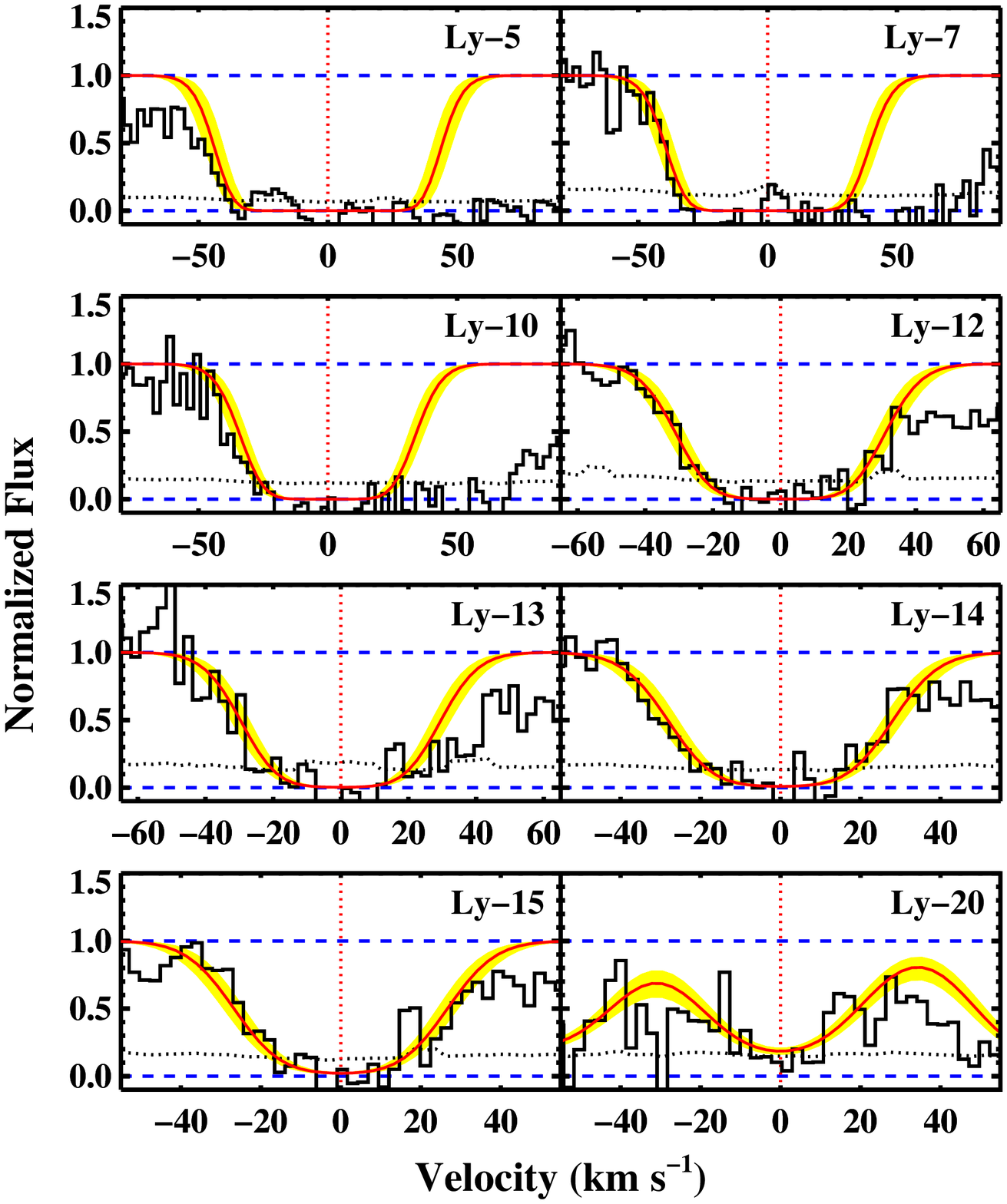}
\end{center}
\end{figure*}
{\bf Figure~\figlymant:}
{\small Same as Figure \figlymano, but for \llstb.}

\clearpage
\begin{figure*}[!ht]
\begin{center}
\includegraphics[scale=0.6]{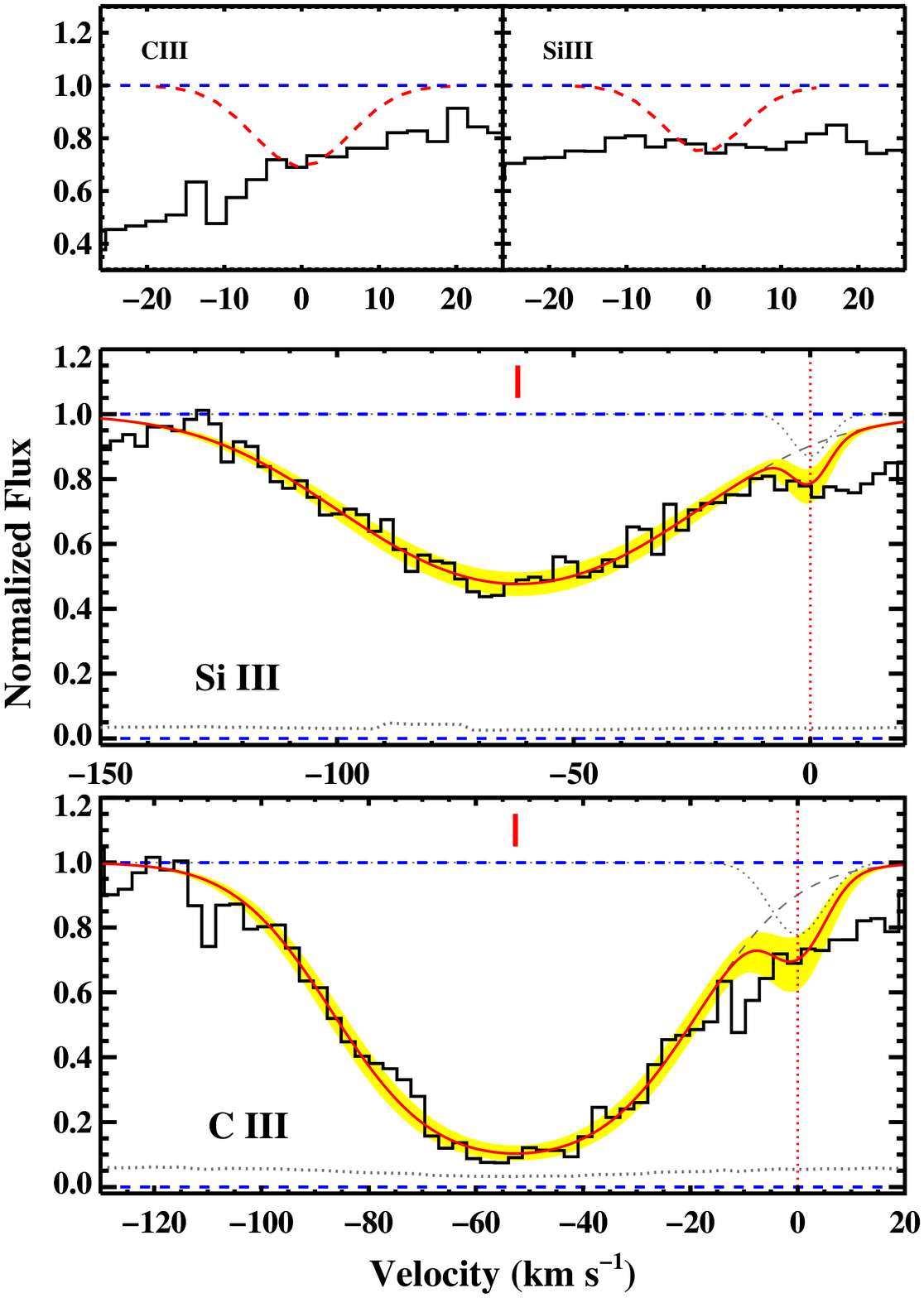}
\end{center}
\end{figure*}
{\bf Figure \figmetline:} 
{\small Top panels:  \ion{C}{III}~977  and \ion{Si}{III}~1206
transitions for \llso. Together with the HIRES data (black histograms),
we display a model for the strongest lines that could be hidden by the IGM 
absorption (red dashed lines). In the bottom panels, we present 
a single component model (gray dashed line) for the IGM absorption 
that contributes to the opacity at the \ion{C}{III}  and \ion{Si}{III} frequencies.
Models for these transitions with the adopted column density limits are 
also shown (gray dotted line). In red, we display the combined two-component models, together
with the corresponding $2\sigma$ errors on the line parameters.
In all cases, we assume a Doppler parameter of $6.9~\rm km~s^{-1}$ for 
carbon and $5.3~\rm km~s^{-1}$ for silicon, as inferred from the hydrogen and 
deuterium lines.}

\clearpage
\begin{figure*}[!ht]
\begin{center}
\includegraphics[scale=0.5]{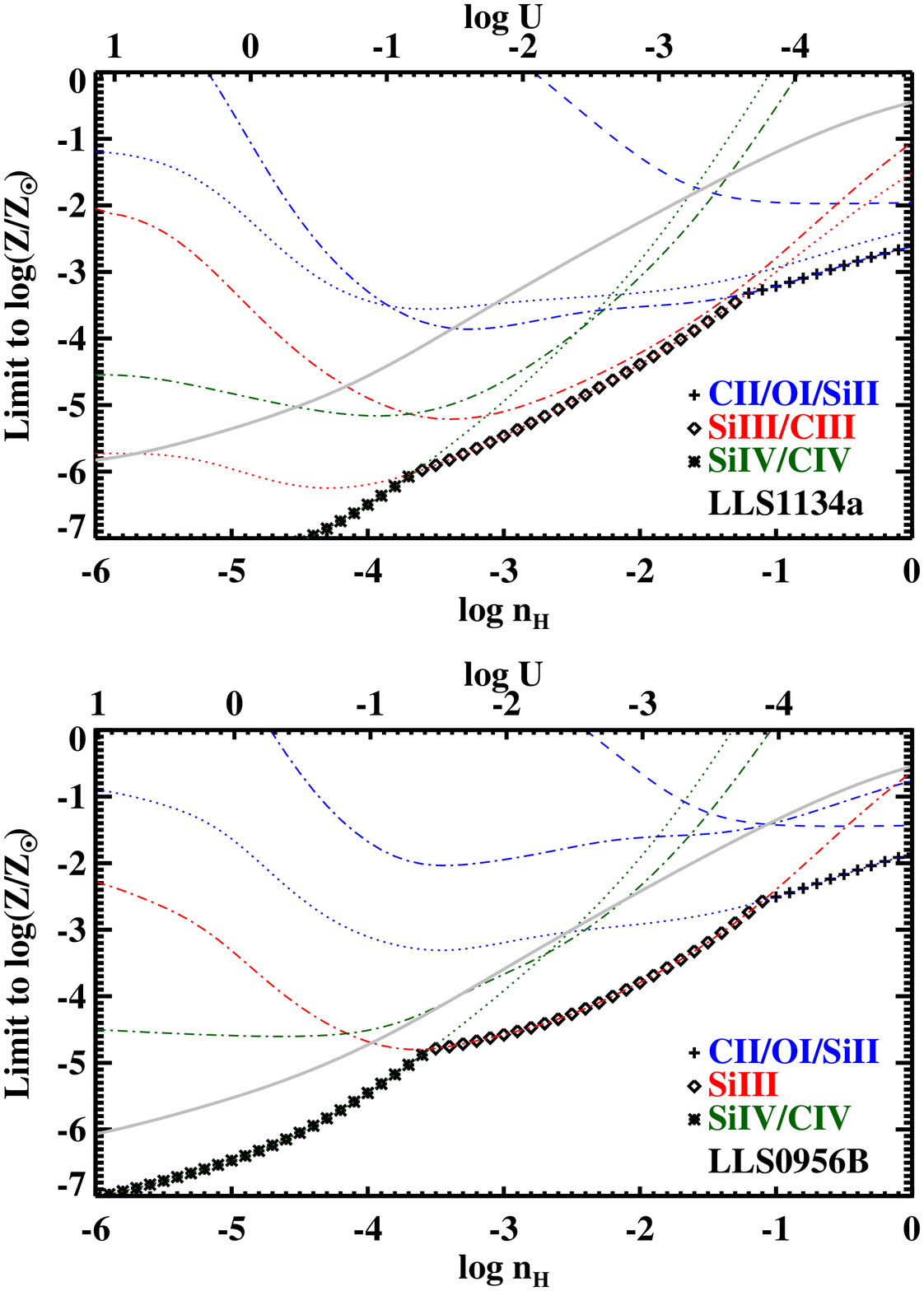}
\end{center}
\end{figure*}
{\bf Figure~\figcla:} 
{\small The two panels describe the metallicity limits to (top) \llso\ and
(bottom) \llstb\ as a function of the gas density $n_{\rm H}$, assuming
the Haardt and Madau EUVB radiation field \cite{haa11}.  Each curve shows the 
metallicity limit for a given ion according to the upper limit on its column density.  
Elements C, O, and Si are traced by dotted, dashed, and dash-dot lines
respectively.   The black
symbols then trace the lowest metallicity imposed by the full set of
measurements at each \nh. The solid gray curves in the figure indicate the logarithm of the
neutral fraction of the gas $x_{\rm HI} \equiv n_{\rm HI}/n_{\rm H}$.}

\clearpage
\begin{figure*}[!ht]
\begin{center}
\includegraphics[scale=0.6,angle=90]{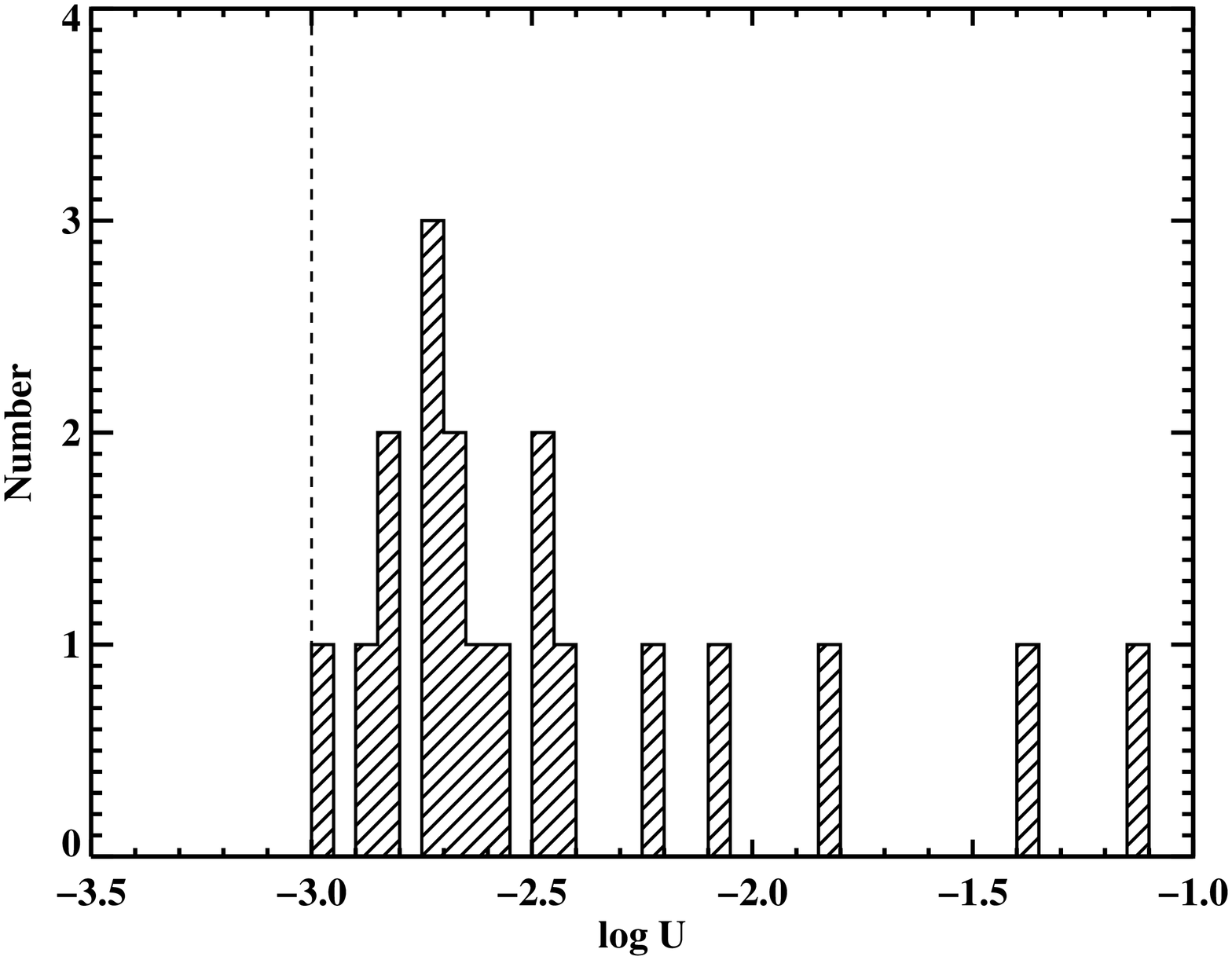}
\end{center}
\end{figure*}
{\bf Figure~\figudis:}
{\small Distribution of the ionization parameter $U$ for all the $z>1.5$ LLSs reported 
to date in the literature. The value $U=10^{-3}$ used in this analysis is marked 
with a dashed line.}

\clearpage
\begin{tabular*}{\textwidth}{@{\extracolsep{\fill}}lcccc}
\hline
\hline
Ion & $z_{\rm{abs}}$  & $\delta v$& Doppler parameter & Column density \\
    &               & ($\rm km~s^{-1}$) &($\rm km~s^{-1}$) & ($\log N$) \\
\hline
\ion{D}{I}&    $3.410883  \pm     0.000004$&  $0$ & $10.2  \pm    0.8$ & $13.26 \pm   0.04$\\
\ion{H}{I}&    $3.40997   \pm     0.00002$ &$-62$ & $ 9.9  \pm    1.9$ & $12.99 \pm   0.07$\\
\ion{H}{I}&    $2.7209    \pm     0.0001$  &  $-$ & $36.9  \pm    9.2$ & $13.05 \pm   0.12$\\
\ion{H}{I}&    $3.410883  \pm     0.000004$&  $0$ & $15.4  \pm    0.3$ & $17.94 \pm   0.05$\\
\ion{H}{I}&    $3.41167   \pm     0.00001$ & $54$ & $18.0  \pm    0.9$ & $16.71 \pm   0.02$\\
\hline
\end{tabular*}
\newline
\newline

\newline
{\bf Table \tabfitllso:} 
{\small Summary of the best-fit model for the LLS1134 hydrogen and deuterium 
absorption. For each ion we list: the redshift of the absorption, the velocity
offset relative to the main hydrogen component, the Doppler parameter and the 
column density.}

\clearpage
\clearpage
\begin{tabular*}{\textwidth}{@{\extracolsep{\fill}}crrcccc}
\hline
\hline
 & & &\multicolumn{2}{c}{\llso}&\multicolumn{2}{c}{\llstb}\\
 Ion&$\lambda$&$\log f$&$\log N$&$W_0~^{a}$&$\log N$&$W_0~^{a}$\\
    &(\AA)    &        &  ($\rm cm^{-2}$)  &(m\AA)  & ($\rm cm^{-2}$) &(m\AA)\\
\hline
\ion{C}{II}  &1334.5323 &$ -0.8935$& $< 12.45$& $<5.7$ & $<12.26$ & $<3.7$  \\
\ion{C}{III} & 977.0200 &$ -0.1180$& $< 12.20$&$<10.2$ & -        & -       \\
\ion{C}{IV}  &1548.1950 &$ -0.7194$& $< 12.09$& $<5.0$ & $<12.30$ & $<8.1$  \\
\ion{O}{I}   &1302.1685 &$ -1.3110$& $< 12.76$& $<4.2$ & $<12.50$ & $<2.3$  \\
\ion{Si}{II} &1260.4221 &$  0.0030$& $< 11.30$& $<2.8$ & -	  & -       \\
\ion{Si}{II} &1526.7066 &$ -0.8962$& $< 12.27$& $<4.9$ & $<12.45$ & $<7.4$  \\
\ion{Si}{III}&1206.5000 &$  0.2201$& $< 11.40$& $<5.4$ & $<11.21$ & $<3.5$  \\
\ion{Si}{IV} &1393.7550 &$ -0.2774$& $< 11.69$& $<4.4$ & $<11.83$ & $<6.1$  \\
\ion{Fe}{II} &1608.4511 &$ -1.2366$& $< 12.59$& $<5.2$ & $<12.72$ & $<7.0$  \\
\hline
\end{tabular*}
{\footnotesize 
$^a$ The rest-frame equivalent width ($W_0$) is computed for the linear portion of the
curve of growth.}
\newline
\newline

\newline
{\bf Table \tabmetlim:} 
{\small $2\sigma$ upper limits on the column density 
of metal ions and rest-frame equivalent widths. 
Rest frame wavelengths and oscillator strengths ($f$) from \cite{mor03} are also listed.}

\clearpage
\clearpage
\begin{tabular*}{\textwidth}{@{\extracolsep{\fill}}lcccc}
\hline
\hline
Quasar&$z_{\rm{abs}}$&$\log~\rm D/H$&$\log ~\rm N_{HI}$&$\rm \left[X/H\right]$\\
\hline
HS010$5+1$619    & 2.53600 & $-4.60\pm 0.04$  & $19.42 \pm 0.01$ & $-2.00^a$\\
Q091$3+07$2      & 2.61843 & $-4.56\pm 0.04$  & $20.34 \pm 0.04$ & $-2.37^a$\\
Q100$9+2$99      & 2.50357 & $-4.40\pm 0.07$  & $17.39 \pm 0.06$ & $-2.5^b$\\
\nameone         &\tabzone & \dhone           & \nhione          & \metone$^{b,c}$\\
Q124$3+3$047     & 2.52566 & $-4.62\pm 0.05$  & $19.73 \pm 0.04$ & $-2.79^a$\\
SDSSJ155$8-0$031 & 2.70262 & $-4.48\pm 0.06$  & $20.67 \pm 0.05$ & $-1.47^a$\\
Q193$7-1$009     & 3.57220 & $-4.48\pm 0.04$  & $17.86 \pm 0.02$ & $-2.7,-1.9^{b,d}$\\
Q220$6-1$99      & 2.07624 & $-4.78\pm 0.09$  & $20.43 \pm 0.04$ & $-2.04^a$\\
\hline
\end{tabular*}
{\footnotesize 
$^a$ Metallicity obtained using oxygen.
$^b$ Metallicity obtained using silicon.
$^c$ This work.
$^d$ The authors report metallicities for two components.
Data from \cite{ome06,pet08b,kir03,ome01,bur98a,bur98b,pet01,pet08a}. 
}
\newline
\newline

{\bf Table \tabdeutsys:} 
{\small Deuterium abundances in $z>2$ absorption line systems.}



\end{document}